\documentclass[letterpaper]{article} % DO NOT CHANGE THIS
\usepackage{aaai25}  % DO NOT CHANGE THIS
\usepackage{times}  % DO NOT CHANGE THIS
\usepackage{helvet}  % DO NOT CHANGE THIS
\usepackage{courier}  % DO NOT CHANGE THIS
\usepackage[hyphens]{url}  % DO NOT CHANGE THIS
\usepackage{graphicx} % DO NOT CHANGE THIS
\def\UrlFont{\rm}  % DO NOT CHANGE THIS
\usepackage{natbib}  % DO NOT CHANGE THIS AND DO NOT ADD ANY OPTIONS TO IT
\usepackage{caption} % DO NOT CHANGE THIS AND DO NOT ADD ANY OPTIONS TO IT
\usepackage{algorithm}
\usepackage{algorithmic}

\usepackage{newfloat}
\usepackage{listings}
\DeclareCaptionStyle{ruled}{labelfont=normalfont,labelsep=colon,strut=off} % DO NOT CHANGE THIS
\floatstyle{ruled}
\newfloat{listing}{tb}{lst}{}
\floatname{listing}{Listing}
\title{Approximate State Abstraction for Markov Games}
\author{
    Hiroki Ishibashi\textsuperscript{\rm 1}, Kenshi Abe\textsuperscript{\rm 1, 2}, Atsushi Iwasaki\textsuperscript{\rm 1}
}
\affiliations{
    \textsuperscript{\rm 1}The University of Electro-Communications\\
    \textsuperscript{\rm 2}CyberAgent\\
}

\usepackage{amsmath}
\usepackage{amsthm}
\usepackage{comment}
\usepackage{subcaption}
\newtheorem{theorem}{Theorem}

\newtheorem{lemma}{Lemma}

\newtheorem{assumption}{Assumption}

\newcommand{\argmax}{\mathop{\rm arg~max}\limits}
\newcommand{\argmin}{\mathop{\rm arg~min}\limits}

\begin{document}

\maketitle

\begin{abstract}
This paper introduces state abstraction for two-player zero-sum Markov games (TZMGs), where the payoffs for the two players are determined by the state representing the environment and their respective actions, with state transitions following Markov decision processes%
% (MDPs)
. 
For example, in games like soccer, the value of actions changes according to the state of play, and thus such games should be described as Markov games. 
In TZMGs, as the number of states increases, computing equilibria becomes more difficult. 
Therefore, we consider state abstraction, which reduces the number of states by treating multiple different states as a single state.
There is a substantial body of research on finding optimal policies for Markov decision processes using state abstraction. 
However, in the multi-player setting, the game with state abstraction may yield different equilibrium solutions from those of the ground game. 
To evaluate the equilibrium solutions of the game with state abstraction, we derived bounds on the duality gap, which represents the distance from the equilibrium solutions of the ground game. 
Finally, we demonstrate our state abstraction with Markov Soccer, compute equilibrium policies, and examine the results.
\end{abstract}


\section{Introduction}\label{sec:introduction}

% MARL has achieved great success
Multi-agent reinforcement learning (MARL) is a framework for sequential decision-making, where multiple agents make decisions in a non-stationary environment to maximize their cumulative rewards. 
MARL has a wide range of applications, e.g., robotics, distributed control, game AI, and so on~\cite{shalev::2016,silver2016mastering,silver2017mastering,brown:science:2018,perolat:science:2022}. 
%
% MG is solvable but suffers from the exponential growth of state space size
Such an environment is often modeled as two-player zero-sum Markov games (TZMGs)~\cite{littman1994markov} and computing the equilibria is said to be empirically tractable. However, it still suffers from the exponential growth of state space size in the number of domain variables. 

% MDP faces the same challenge, and some state abstraction techniques have been developed. 
Markov decision processes (MDPs), which are a single-agent version of Markov games, face the same challenge. 
Solving MDPs, i.e., computing an optimal policy, is P-Complete in state space size~\cite{littman1994markov}, while that size often exponentially increases. 
% Exact state abstraction vs. Approximate state abstraction
Several state abstraction techniques have been developed, which aggregate multiple states into one abstract state according to a certain criterion and reduce the state space size. For example, a state is equivalent to another state if choosing an action leads to the same state with same rewards.
Such abstraction is effective for agents to solve more complicated MDPs than they would be able to without using abstraction. However, if we abstract the state space by regarding only identical states as equivalent, we are difficult to find all of them within a reasonable time, and even worse no two states might be identical. 

% We take an approximate state abstraction approach
%% We extend the work of AHL for single-agent MDP to two-player zero-sum Markov games
In contrast to such \textit{exact state abstraction}, \citet{abel2016near} proposed \textit{approximate state abstraction}, which characterizes how close a state is to another state in single-agent MDPs. This technique reduces ground MDPs with large state spaces to abstract MDPs with smaller state spaces by aggregating states according to some notion of closeness or similarity. While this relaxation makes the state spaces smaller, the resulting optimal policies in abstract MDPs may become suboptimal. Furthermore, they derive error bounds for the resulting policies based on four different criteria, such as optimal Q-values, according to which states are aggregated.

TZMGs extend MDPs to model decision-making by two interacting agents in a shared, state-dependent environment. Unlike MDPs, the rewards in TZMGs depend on the actions of both agents. For instance, in a soccer game, the rewards vary based on the state of play and the actions chosen by both players. TZMGs provide a framework to formalize such scenarios and have stimulated research in competitive MARL. An agent's optimal policy depends on the policy of its opponent. The goal is to identify the equilibrium policy profile, which consists of mutual best responses, to predict and understand the consequences of their interactions.

Building upon these insights, this paper extends the work by \citet{abel2016near} from single-agent MDPs to TZMGs. We first describe an approximate state abstraction based on optimal Q-value criteria or minimax values and derive the bound of the duality gap for the resulting equilibrium, which measures the proximity to equilibrium. To establish the bound, we analyze the gains from best responses in the ground game -- where the strategy space is unabstracted -- against the resulting equilibrium in the abstracted game, where the strategy space has been simplified. Second, we conduct experiments in Markov Soccer and demonstrate how state spaces are reduced and how well the equilibrium strategies in the ground TZMG are approximated. Furthermore, we discuss the extension for the other three criteria. % mentioned in \citet{abel2016near}.

% Building upon those insights, this paper extends the work by \citet{abel2016near} from single-agent MDPs to TZMGs. We first describe the optimal Q-value criteria or minimax values and derive the bound of the duality gap, which measures the proximity to equilibrium for a given policy profile. 
% Second, we provide some experiments in Markov Soccer as a representative TZMGs and demonstrate how state spaces is reduced and how much duality gaps are approximated. Furthermore, we discuss the extension for the other three criteria raised in \cite{abel2016near}. 

\paragraph{Related Literature.}
% Two-player zero-sum Markov games
% \subsection{Policy Optimization for Markov Games}
% \begin{itemize}
%     \item マルコフゲームにおける均衡解を求める有名なアルゴリズムの一つとして Minimax Q-learning~\cite{littman1994markov}がある．
%     \item 強化学習における Q-learning と，ゲーム理論におけるミニマックス戦略を組み合わせたアルゴリズムである．本研究ではこのアルゴリズムを用いて均衡計算を行う．
%     \item 一般和マルコフゲームの均衡解を求めるアルゴリズムには Nash Q-learning がある．
%     \item ベースとしているアルゴリズムは Q-learning であり，状態価値関数の更新時にナッシュ均衡時の状態価値関数を用いる．
%     \item Minimax Q-learning と Nash Q-learning はテーブル形式のアルゴリズムのため，状態数が有限のゲームのみで適用可能である．
%     \item Nash-DQN は，関数近似により状態数が無限のゲームに対する均衡計算が可能である．
% \end{itemize}

% State abstraction MDP
% \subsection{State Abstraction for Markov Decision Process}

There is a lot of literature on state abstraction for MDPs initiated by \cite{dietterich1998maxq,dietterich1999state,jonsson2000automated}. 
The concept of state equivalence has been developed to reduce state space size by aggregating equivalent states~\cite{givan2003equivalence}, and it has been relaxed by allowing some associated actions~\cite{ravindran2003smdp,ravindran2004approximate,van2020plannable}. 
\citet{ferns2004metrics} proposed a distance between states and aggregates states with zero distance.
Similarly, \citet{castro2020scalable} parameterized the \textit{bisimulation} metric~\cite{ferns2004metrics}. 
Recently, \citet{dadvar2023conditional} incorporated state abstraction into policy iteration. 
\citet{li2006towards} discussed the optimality of policies on five criteria for state abstraction. 

In another line of work, state space abstraction has been developed in poker AI~\cite{gilpin2006competitive,gilpin2006finding,gilpin2007better,gilpin2007potential,johanson2013evaluating,waugh2013fast,ganzfried2013action,burch2014solving,kroer2018unified}.
The state spaces in poker can have at most $10 \times 10^{160}$ states, and this area has been extensively investigated. 
Initially, states were abstracted manually based on knowledge and experience in poker. 
Subsequently, automated abstraction techniques were developed.
For example, states are aggregated by estimating winning probabilities at information sets~\cite{gilpin2007better,gilpin2007potential,johanson2013evaluating,waugh2013fast,ganzfried2013action}. 
Furthermore, \citet{kroer2018unified} presented a unified framework for analyzing abstractions that can express all types of abstractions and solution concepts used in prior work, with performance guarantees, while maintaining comparable bounds on abstraction quality.

\section{Preliminaries}
\subsection{Two-Player Zero-Sum Markov Games}
\label{subsection:TZMGs}
A two-player zero-sum Markov game (TZMG) $\mathcal{M}$ is defined by a tuple $\mathcal{M} = \langle \mathcal{S},\mathcal{A}_1,\mathcal{A}_2,P,R,\gamma\rangle$.
Here, $\mathcal{S}$ represents a finite state space, $\mathcal{A}_i$ represents an action space for player $i\in \{1, 2\}$, $P:\mathcal{S}\times\mathcal{A}_1\times\mathcal{A}_2 \rightarrow \Delta(\mathcal{S})$ represents a transition probability function, $R:\mathcal{S}\times\mathcal{A}_1\times\mathcal{A}_2\rightarrow [0,1]$ represents a reward function, and $\gamma\in[0,1)$ represents a discount factor.
Let $\mathcal{A} = \mathcal{A}_1 \times \mathcal{A}_2$, and let $\boldsymbol{a} = (a_1, a_2) \in \mathcal{A}$ denote the action profile.
For a given state $s\in \mathcal{S}$ and an action profile $\boldsymbol{a}\in \mathcal{A}$, the next state is determined according to $P(\cdot | s, \boldsymbol{a})$, and player 1 (resp. player 2) receives a reward of $R(s, \boldsymbol{a})$ (resp. $-R(s, \boldsymbol{a})$).

A Markov policy for player $i$, denoted as $\pi_i: \mathcal{S} \to \Delta(\mathcal{A}_i)$, represents the probability of choosing action $a_i\in \mathcal{A}_i$ at a state $s\in\mathcal{S}$.
Letting $\boldsymbol{\boldsymbol{\pi}} = (\pi_1, \pi_2)$ be a policy profile, we further define the state value function, which is the expected discounted sum of rewards at state $s\in\mathcal{S}$ as follows:
\begin{align*}
    V^{\boldsymbol{\pi}}(s) := \mathrm{E} \Biggl[&\sum_{t=1}^{\infty}\gamma^{t-1} R(s_t,\boldsymbol{a}_t) ~\bigg|~ s_1 = s, \\
    &\boldsymbol{a}_t\sim\boldsymbol{\pi}(\cdot|s_t),s_{t+1}\sim P(\cdot | s_t,\boldsymbol{a}_t), \forall t\geq0 \Biggr].
\end{align*}
We similarly define the state-action value function of taking an action profile $\boldsymbol{a}\in\mathcal{A}$ at state $s\in\mathcal{S}$ as follows:
\begin{align*}
    Q^{\boldsymbol{\pi}}(s, \boldsymbol{a}) := R(s, \boldsymbol{a}) + \gamma \sum_{s'\in \mathcal{S}} P(s' | s, \boldsymbol{a}) V^{\boldsymbol{\pi}}(s').
\end{align*}
From the definition of these functions, the state value function $V^{\boldsymbol{\pi}}$ can be expressed as follows:
\begin{align*}
    V^{\boldsymbol{\pi}}(s) = \sum_{\boldsymbol{a}\in\mathcal{A}}\boldsymbol{\pi}(\boldsymbol{a}|s)Q^{\boldsymbol{\pi}}(s,\boldsymbol{a}).
\end{align*}

\subsection{Nash Equilibrium}
In TZMGs, a \textit{Nash equilibrium} is defined as the policy profile that satisfies the following condition for all $s\in \mathcal{S}$ simultaneously:
\begin{align*}
    &\forall (\pi_1, \pi_2), ~V^{\pi_1^\ast,\pi_2}(s) \geq V^{{\boldsymbol{\pi}}^\ast}(s) \geq V^{\pi_1, \pi_2^{\ast}}(s).
\end{align*}
Intuitively, a Nash equilibrium is the policy profile where no player can improve her value by deviating from her own policy.
\citet{shapley1953stochastic} has shown that any Nash equilibrium $\boldsymbol{\pi}^{\ast}$ in TZMGs satisfies the following condition for all $s\in \mathcal{S}$:
\begin{align}
\begin{aligned}
    V^{{\boldsymbol{\pi}}^{\ast}}(s) &= \max_{p\in\Delta(\mathcal{A}_1)}\min_{a_2\in\mathcal{A}_2} \sum_{a_1\in\mathcal{A}_1}p(a_1)Q^{{\boldsymbol{\pi}}^\ast}(s,\boldsymbol{a}) \\
    &= \min_{p\in\Delta(\mathcal{A}_2)}\max_{a_1\in\mathcal{A}_1} \sum_{a_2\in\mathcal{A}_2}p(a_2)Q^{{\boldsymbol{\pi}}^\ast}(s,\boldsymbol{a}).
\end{aligned}
\label{eq:shapley_theorem}
\end{align}
It is known that this minimax value is unique for each $s$ \cite{shapley1953stochastic}, thus we can write $V^{\ast}(s) := V^{\boldsymbol{\pi^{\ast}}}(s)$ and $Q^{\ast}(s, \boldsymbol{a}):= Q^{\boldsymbol{\pi^{\ast}}}(s, \boldsymbol{a})$.

To measure the proximity to equilibrium for a given policy profile ${\boldsymbol{\pi}}=(\pi_1,\pi_2)$, we use the \textit{duality gap} defined as follows:
\begin{align*}
    \mathrm{GAP}\left({\boldsymbol{\pi}}\right) := \max_{s\in \mathcal{S}, \pi_1', \pi_2'}\left(V^{\pi_1', \pi_2}(s) - V^{\pi_1, \pi_2'}(s)\right).
\end{align*}
From the definition, we can see that $\mathrm{GAP}\left({\boldsymbol{\pi}}\right)\geq 0$ for any $\boldsymbol{\pi}$, and the equality holds if and only if $\boldsymbol{\pi}$ is a Nash equilibrium.

\subsection{Minimax Q-learning}
Let us briefly describe Minimax Q-learning~\cite{littman1994markov} which is developed to address the limitations of standard Q-learning in adversarial, zero-sum environments. Its robustness and adaptability in competitive settings make it effective for AI in games and security applications. 
It has been shown that Minimax Q-learning converges to a Nash equilibrium under some appropriate conditions \cite{szepesvari1999unified}.
Due to its theoretical guarantee and ease of implementation, Minimax Q-learning is used as a standard algorithm to compute equilibrium policies for TZMGs. 

Algorithm~\ref{algorithm:minimax-Q learning} illustrates the procedure with finite $T$ iterations. 
\begin{enumerate}
\item At each iteration $t\geq 0$, each player $i\in \{1, 2\}$ chooses an actions $a_{i,t}\in \mathcal{A}_i$ randomly with probability $\beta$, otherwise based on her policy $\pi_{i}^t(\cdot|s_t)\in \Delta(\mathcal{A}_i)$. 
\item State $s_{t}$ transits to $s_{t+1}$ according to the transition probability function $P(\cdot\mid s_t, a_{1,t}, a_{2,t})$.  
\item State-action value function $Q_t$ is updated with the obtained reward $R(s_t,\boldsymbol{a}_t)$ and learning rate $\alpha_t$: 
\begin{align*}
    Q_{t+1}(s_t, \boldsymbol{a}_t)
    :=& (1 - \alpha_t)Q_t(s_t, \boldsymbol{a}_t) \\
    &+ \alpha_t( R(s_t,\boldsymbol{a}_t) + V_t(s_{t+1})).
\end{align*}
\item Players update their policies using linear programming: 
\begin{align*}
    \pi_{1}(\cdot|s_t) := \argmax_{p \in \Delta(\mathcal{A}_1)}\min_{a_{2} \in \mathcal{A}_{2}} \sum_{a_{1}\in A_1}p(a_{1})Q_{t+1}(s_t,\boldsymbol{a}), \\
    \pi_{2}(\cdot|s_t) := \argmin_{p \in \Delta(\mathcal{A}_2)}\max_{a_{1} \in \mathcal{A}_{1}} \sum_{a_{2}\in A_2}p(a_{2})Q_{t+1}(s_t,\boldsymbol{a}).
\end{align*}
\item They update their state value function with current policy:
\begin{align*}
    V_{t+1}(s_t) := \sum_{a_{1}\in \mathcal{A}_1}\sum_{a_{2}\in \mathcal{A}_2}\pi_1(a_{1}|s_t)\pi_2(a_{2}|s_t)Q_{t+1}(s_t,\boldsymbol{a}).
\end{align*}
\end{enumerate}

We implement this procedure when we compute the duality gap in Section~\ref{sec:experiments}. Note that our proposed state abstraction does not depend on Minimax Q-learning as well as in \cite{abel2016near}. In fact, we are going to discuss the extensions to different criteria of aggregating states.

% \resizebox{0.5\linewidth}{!}{
\begin{algorithm}[tb]
    \caption{Minimax Q-learning}
    \label{algorithm:minimax-Q learning}
    \begin{algorithmic}[1]
    \REQUIRE{Learning rates $\alpha_t$, exploration parameter $\beta$}
    \STATE $V[s] \gets 0$ for all $s\in \mathcal{S}$
    \STATE $Q[s, \boldsymbol{a}] \gets 0$ for all $s\in \mathcal{S}$ and $\boldsymbol{a}\in \mathcal{A}$
    \STATE $\pi_i^0(\cdot|s) \gets\left(\frac{1}{|\mathcal{A}_i|}\right)_{a\in \mathcal{A}_i}$ for all $i\in \{1, 2\}$ and $s\in \mathcal{S}$
    \STATE Sample initial state $s_0$
    \FOR{$t=0, 1, \dots, T-1$}
        \STATE $\pi_i'(\cdot | s_t) \gets (1 - \beta) \pi_i^t(\cdot | s_t) + \frac{\beta}{|\mathcal{A}_i|}\mathbf{1}$ for all $i\in \{1, 2\}$
        \STATE Sample action profile $\boldsymbol{a}_t\sim \boldsymbol{\pi}'(\cdot | s_t)$
        \STATE Next state is sampled $s_{t+1}\sim P(\cdot | s_t, \boldsymbol{a}_t)$
        \STATE $Q[s_t, \boldsymbol{a}_t]\gets (1 - \alpha_t)Q[s_t, \boldsymbol{a}_t] $ \\ $\phantom{========}+ \alpha_t\left(R(s_t,\boldsymbol{a}_t) + \gamma V^{{\boldsymbol{\pi}}}[s_{t+1}]\right)$
        \STATE $\pi_1^{t+1}(\cdot|s_t) \!\gets \!\argmax_{p\in\Delta(\mathcal{A}_1)}\min\limits_{a_2\in\mathcal{A}_2} \sum_{a_1\in\mathcal{A}_1}p(a_1)Q[s_t,\boldsymbol{a}_t]$
        \STATE $\pi_2^{t+1}(\cdot|s_t) \!\gets \!\argmin_{p\in\Delta(\mathcal{A}_2)}\max\limits_{a_1\in\mathcal{A}_1} \sum_{a_2\in\mathcal{A}_2}p(a_2)Q[s_t,\boldsymbol{a}_t]$
        \STATE $V(s_t) \!\gets \! \sum\limits_{a_1\in\mathcal{A}_1}\sum\limits_{a_2\in\mathcal{A}_2} \! \pi_1^{t+1}(a_1|s_t)\pi_2^{t+1}(a_2|s_t)Q[s_t,\boldsymbol{a}_t]$
    \ENDFOR
    \ENSURE{$(\pi_1^T, \pi_2^T)$}
    \end{algorithmic}
\end{algorithm}
% }

\section{State Abstraction}
In this section, we extend state abstraction for a MDP to a TZMG.
State abstraction is a method for reducing the state space by aggregating similar states to decrease the time of calculating the equilibria.
In the previous research of Abel et al. \cite{abel2016near}, they propose four different state abstraction approaches for a MDP and theoretically analyze how well the optimal policy in the abstract MDP achieves performance in the ground MDP using various metrics.
In this section, we extend their approach based on state-action value function similarity.
The other approaches, including model similarity, Boltzmann distribution similarity, and multinomial distribution similarity, are introduced in Discussion section.

\subsection{Abstract Two-Player Zero-Sum Markov Games}
This section defines an abstract TZMG $\mathcal{M}_A = \langle \mathcal{S}_A,\mathcal{A}_1,\mathcal{A}_2,P_A,R_A,\gamma\rangle$ by using the notation introduced by \citet{abel2016near,li2006towards}.
Here, $\mathcal{S}_A$ is an abstract state space, and $P_A: \mathcal{S}_A\times \mathcal{A}_1\times \mathcal{A}_2\to \Delta(\mathcal{S}_A)$ is an abstract transition probability function, and $R_A: \mathcal{S}_A\times \mathcal{A}_1 \times \mathcal{A}_2 \to [0, 1]$ is an abstract reward function.

Let $\phi: \mathcal{S} \to \mathcal{S}_A$ be a {\it state aggregation function} that maps from the ground state space $\mathcal{S}$ to an abstract state space $\mathcal{S}_A$.
Given $\phi$, we can define a set of states $G_A(s_A)$ that are aggregated into the same abstract state $s_A\in \mathcal{S}_A$:
\begin{align*}
    G_A(s_A) := \{g \in \mathcal{S} \mid \phi(g) = s_A\}.
\end{align*}
For a given ground state $s\in \mathcal{S}$, we further define a set of states $G(s)$ that are aggregated into the same abstract state as $s$:
\begin{align*}
    G(s) := \{g \in \mathcal{S} \mid \phi(g) = \phi(s)\}.
\end{align*}

In order to construct an abstract transition probability function $P_A$ and an abstract reward function $R_A$, we introduce a {\it weight function} $w: \mathcal{S}\to [0, 1]$, which satisfies the following condition:
\begin{align*}
    \forall s_A\in \mathcal{S}_A, ~\sum_{g\in G_A(s_A)} w(g) = 1.
\end{align*}
For example, $w(s) = 1/|G(s)|$ is a representative weight function.
Using this function, we define the abstract transition probability function $P_A$ as follows:
\begin{align*}
    P_A(s_A' | s_A, \boldsymbol{a}) := \sum_{s\in G_A(s_A)} \sum_{s'\in G_A(s'_A)} P(s'|s,\boldsymbol{a}) w(s).
\end{align*}
Similarly, the abstract reward function $R_A: \mathcal{S}_A \times \mathcal{A} \to [0,1]$ is defined as follows:
\begin{align*}
    R_A(s_A,\boldsymbol{a}) := \sum_{s\in G_A(s_A)} R(s,\boldsymbol{a}) w(s).
\end{align*}

The policy profile in the abstract TZMG $\mathcal{M}_A$ is defined in a similar manner to the ground TZMG $\mathcal{M}$.
Let $\pi_{A,i}: \mathcal{S}_A \to \Delta(\mathcal{A}_i)$ be a policy for player $i$ in $\mathcal{M}_A$.
Similarly, $V_A^{\boldsymbol{\pi}_A}(s_A)$ denotes a state value for a given policy profile ${\boldsymbol{\pi}}_A$ at state $s_A\in \mathcal{S}_A$ in the abstract TZMG, and $Q_A^{\boldsymbol{\pi}_A}(s_A, \boldsymbol{a})$ is defined as a state-action value for $s_A\in \mathcal{S}_A$ and $\boldsymbol{a}\in \mathcal{A}$.
Finally, letting $\boldsymbol{\pi}_A^{\ast}$ be a Nash equilibrium in the abstract TZMG $\mathcal{M}_A$, $\boldsymbol{\pi}_A^{\ast}$ must satisfy the following condition for all $s_A\in \mathcal{S}_A$ and $(\pi_{A,1}, \pi_{A,2})$:
\begin{align*}
    &V_{A}^{\pi_{A,1}^\ast,\pi_{A,2}}(s_A) \geq V_{A}^{{\boldsymbol{\pi}}_A^\ast}(s_A) \geq V_{A}^{\pi_{A,1}, \pi_{A,2}^{\ast}}(s_A).
\end{align*}

\section{Abstraction Based on Minimax Values}
In this section, we analyze the performance of a Nash equilibrium $\boldsymbol{\pi}_A^{\ast}$ in an abstract TZMG $\mathcal{M}_A$ under a specific aggregation function $\phi$ when applied to the ground TZMG $\mathcal{M}$.
To this end, we define the policy profile ${\boldsymbol{\pi}}_{GA}^{\ast}(s)$ in the ground TZMG, which is induced by $\boldsymbol{\pi}_{A}^{\ast}$.
Formally, $\boldsymbol{\pi}_{GA}^{\ast}(s)$ for all $s$ in $\mathcal{S}$ is given by:
\begin{align*}
    \boldsymbol{\pi}_{GA}^{\ast}(s) := \boldsymbol{\pi}_{A}^{\ast}(\phi(s)).
\end{align*}
In a later section, we derive the upper bound on the duality gap of $\boldsymbol{\pi}_{GA}^{\ast}$ under various aggregation functions $\phi$.

\subsection{Suboptimality of Nash Equilibria in Abstract TZMGs}
We examine the aggregation function $\phi^{Q^{\ast}}$, which is constructed based on the state-action value function $Q^{\ast}$.
Specifically, in the abstraction $\phi^{Q^{\ast}}$, states are aggregated to the same abstract state when their minimax state-action values are close within $\epsilon$.
\begin{assumption}
\label{asmp:minimax_Q_abstraction}
The apggregation function $\phi^{Q^{\ast}}$ satisfies the following property for some non-negative constant $\epsilon\geq 0$:
 {\small
\begin{align}
    & \phi^{Q^{\ast}}(s_1) = \phi^{Q^{\ast}}(s_2) \nonumber \\
    & \Rightarrow  \forall \boldsymbol{a}\in \mathcal{A}, ~  \left|Q^{\ast}(s_1, \boldsymbol{a}) - Q^{\ast}(s_2, \boldsymbol{a})\right| \leq \epsilon.
    \label{eq:minimax_value_assumption}
\end{align}}
\end{assumption}

Under Assumption \ref{asmp:minimax_Q_abstraction}, it can be shown that the suboptimality of $\boldsymbol{\pi}_{GA}^{\ast}$ is no more than $\mathcal{O}(\epsilon)$.
\begin{theorem}
\label{thm:error_wrt_minimax_value}
When the ground states are aggregated by the apggregation function $\phi^{Q^{\ast}}$ satisfying Assumption \ref{asmp:minimax_Q_abstraction} with $\epsilon \geq 0$, then $\boldsymbol{\pi}_{GA}^{\ast}$ satisfies:
 {\small
 \begin{align*}
    \mathrm{GAP}\left(\boldsymbol{\pi}_{GA}^{\ast}\right) \leq \frac{12\epsilon}{(1-\gamma)^3}.
\end{align*}}
\end{theorem}

To perform an initial abstraction, minimax Q-learning is used to calculate Q-values for the ground game. If the results do not meet a sufficient solution criterion, the process iteratively updates the Q-values and refines the abstraction~\cite{li2006towards}. Developing efficient algorithms for discovering abstractions remains open. Also, such abstractions have potential utility beyond a single game, enabling transferable knowledge across related games~\cite{jong:ijcai:2005}.

% To perfrom an initial abstraction, we need to run minimax Q-learning on the ground game to obtain Q-values after a predefined number of iterations. If this approximation does not meet a sufficient solution criterion, we can iterate by updating Q-values and refining the abstraction. While we believe such iterative refinement holds promise, developing efficient discovery algorithms for certain abstractions remains an open question, as highlighted by \citet{li2006towards}. This exploration, we believe, could lead to practical advancements in state abstraction. We would also like to point out that identifying abstractions has potential utility beyond a single game, as they can facilitate solving families of related Markov games. Specifically, such abstractions can function as transferable knowledge across similar tasks, as suggested by \citet{jong:ijcai:2005,li2006towards}

\subsection{Proofs for Theorem \ref{thm:error_wrt_minimax_value}}
This section provides the proofs for Theorem \ref{thm:error_wrt_minimax_value}.

\paragraph{Proof of Theorem \ref{thm:error_wrt_minimax_value}.}
By the definition of the duality gap:
 {\small
\begin{align}
    \mathrm{GAP}\left({\boldsymbol{\pi}}_{GA}^{\ast}\right) = & \max_{s\in \mathcal{S}, \pi_1, \pi_2}\left(V^{\pi_1, \pi_{GA,2}^{\ast}}(s) - V^{\pi_{GA,1}^{\ast}, \pi_2}(s)\right) \nonumber\\
    \leq & \max_{s\in \mathcal{S}, \pi_1}\left(V^{\pi_1, \pi_{GA,2}^{\ast}}(s) - V^{\boldsymbol{\pi}_{GA}^{\ast}}(s)\right) \nonumber\\
    &+ \max_{s\in \mathcal{S}, \pi_2}\left(V^{\boldsymbol{\pi}_{GA}^{\ast}}(s) - V^{\pi_{GA,1}^{\ast}, \pi_2}(s)\right).
    \label{eq:exploitability_of_pi_GA}
\end{align}}
Hence, it is sufficient to derive the upper bound on $\max_{\pi_i} V^{\pi_1, \pi_{GA, 2}^{\ast}}(s) - V^{\boldsymbol{\pi}_{GA}^{\ast}}(s)$ and $V^{\boldsymbol{\pi}_{GA}^{\ast}}(s) - \max_{\pi_i} V^{\pi_{GA, 1}^{\ast}, \pi_2}(s)$ for each $s\in \mathcal{S}$.
Here, letting $\pi_1^{\dagger}$ be an optimal policy against $\pi_{GA, 2}^{\ast}$ in the ground TZMG, we obtain the following result on the performance difference between $\pi_1^{\dagger}$ and $\pi_{GA,1}^{\ast}$ against $\pi_{GA,2}^{\ast}$.
The proof for Lemma \ref{lem:exploitability_bound_minimax_value} is provided in Appendix \ref{app:exploitability_bound_minimax_value}.
\begin{lemma}
\label{lem:exploitability_bound_minimax_value}
Assume that the aggregation function $\phi$ satisfies the following condition for some non-negative constant $\delta \geq 0$:
\begin{align*}
    \forall s\in \mathcal{S}, \boldsymbol{a}\in \mathcal{A}, ~\left|Q_A^{\boldsymbol{\pi}_A^{\ast}}(\phi(s), \boldsymbol{a}) - Q^{\pi_1^{\dagger}, \pi_{GA, 2}^{\ast}}(s, \boldsymbol{a})\right| \leq \delta.
\end{align*}
Then, we have for any $s\in \mathcal{S}$:
 {\small
 \begin{align*}
    V^{\pi_1^{\dagger}, \pi_{GA, 2}^{\ast}}(s) - V^{\boldsymbol{\pi}_{GA}^{\ast}}(s) \leq \frac{2\delta}{1-\gamma}.
\end{align*}}
\end{lemma}

Next, we show that the assumption in Lemma \ref{lem:exploitability_bound_minimax_value} holds with some constant $\delta$.
By the triangle inequality, the difference between $Q_A^{\boldsymbol{\pi}_A^{\ast}}(\phi^{Q^{\ast}}(s), \boldsymbol{a})$ and $Q^{\pi_1^{\dagger}, \pi_{GA, 2}^{\ast}}(s, \boldsymbol{a})$ can be bound as follows:
 {\small
\begin{align}
    &\left|Q_A^{\boldsymbol{\pi}_A^{\ast}}(\phi^{Q^{\ast}}(s), \boldsymbol{a}) - Q^{\pi_1^{\dagger}, \pi_{GA, 2}^{\ast}}(s, \boldsymbol{a})\right| \nonumber \\
    &\leq \left|Q_A^{\boldsymbol{\pi}_A^{\ast}}(\phi^{Q^{\ast}}(s), \boldsymbol{a}) - Q^{\ast}(s, \boldsymbol{a})\right| \nonumber \\ 
    & \phantom{\leq} \qquad + \left|Q^{\ast}(s, \boldsymbol{a}) - Q^{\pi_1^{\dagger}, \pi_{GA, 2}^{\ast}}(s, \boldsymbol{a})\right|.
\label{eq:decomposed_q}
\end{align}}
The first term in \eqref{eq:decomposed_q} means the difference between the minimax state-action values in the abstract TZMG $\mathcal{M}_A$ and ground TZMG $\mathcal{M}$.
The second term in \eqref{eq:decomposed_q} represents the performance gap between $\boldsymbol{\pi}^{\ast}$ and $(\pi_1^{\dagger}, \pi_{GA, 2}^{\ast})$ in $\mathcal{M}$.
Each term can be bounded as follows:
\begin{lemma}
\label{lem:diff_between_q_and_q_abs_wrt_minimax_value}
In the same setup of Theorem \ref{thm:error_wrt_minimax_value}, we have for any $s\in \mathcal{S}$ and $\boldsymbol{a}\in \mathcal{A}$:
 {\small
 \begin{align*}
    \left|Q_A^{\boldsymbol{\pi}_A^{\ast}}(\phi^{Q^{\ast}}(s), \boldsymbol{a}) - Q^{\ast}(s, \boldsymbol{a})\right| \leq \frac{\epsilon}{1-\gamma}.
\end{align*}}

\end{lemma}
\begin{lemma}
\label{lem:diff_between_best_response_q_and_minimax_q}
In the same setup of Theorem \ref{thm:error_wrt_minimax_value}, we have for any $s\in \mathcal{S}$ and $\boldsymbol{a}\in \mathcal{A}$:
 {\small
 \begin{align*}
    \left|Q^{\ast}(s, \boldsymbol{a}) - Q^{\pi_1^{\dagger}, \pi_{GA, 2}^{\ast}}(s, \boldsymbol{a})\right| \leq \frac{2\epsilon}{(1-\gamma)^2}.
\end{align*}}
\end{lemma}

By combining \eqref{eq:decomposed_q} and Lemmas \ref{lem:diff_between_q_and_q_abs_wrt_minimax_value} and  \ref{lem:diff_between_best_response_q_and_minimax_q}, we get for any $s\in \mathcal{S}$ and $\boldsymbol{a}\in \mathcal{A}$:
 {\small
\begin{align*}
    &\left|Q_A^{\boldsymbol{\pi}_A^{\ast}}(\phi^{Q^{\ast}}(s), \boldsymbol{a}) - Q^{\pi_1^{\dagger}, \pi_{GA, 2}^{\ast}}(s, \boldsymbol{a})\right| \leq \frac{3\epsilon}{(1-\gamma)^2}.
\end{align*}}
This inequality implies that, under the aggregation function $\phi^{Q^{\ast}}$, the assumption in Lemma \ref{lem:exploitability_bound_minimax_value} holds with $\delta=\frac{3\epsilon}{(1-\gamma)^2}$.
Thus, we can apply Lemma \ref{lem:exploitability_bound_minimax_value}, and then obtain the following inequality:
 {\small
\begin{align}
    \label{eq:exploitability_bound_wrt_minimax_value}
    V^{\pi_1^{\dagger}, \pi_{GA, 2}^{\ast}}(s) - V^{\boldsymbol{\pi}_{GA}^{\ast}}(s) 
    & \leq \frac{6\epsilon}{(1-\gamma)^3}.
\end{align}}
By a similar procedure, we can show that:
 {\small
\begin{align}
    \label{eq:exploitability_bound_wrt_minimax_value_player2}
    V^{\boldsymbol{\pi}_{GA}^{\ast}}(s) - V^{\pi_{GA, 1}^{\ast}, \pi_2^{\dagger}}(s) & \leq \frac{6\epsilon}{(1-\gamma)^3}.
\end{align}}
By combining \eqref{eq:exploitability_of_pi_GA}, \eqref{eq:exploitability_bound_wrt_minimax_value}, and \eqref{eq:exploitability_bound_wrt_minimax_value_player2}, we can upper bound the duality gap of $\boldsymbol{\pi}_{GA}^\ast$ as follows:
{\small
\begin{align*}
    \mathrm{GAP}\left(\boldsymbol{\pi}_{GA}^{\ast}\right) \leq \frac{6\epsilon}{(1-\gamma)^3}\times 2 = \frac{12\epsilon}{(1-\gamma)^3}. \qed
\end{align*}}

\paragraph{Proof of Lemma \ref{lem:diff_between_q_and_q_abs_wrt_minimax_value}.}
From the definition of the state-action value function in the abstract TZMG $\mathcal{M}_A$:
{\small
\begin{align*}
    &Q_A^{\boldsymbol{\pi}_A^{\ast}}(\phi^{Q^{\ast}}(s), \boldsymbol{a}) \\
    &= R_A(\phi^{Q^{\ast}}(s), \boldsymbol{a}) + \gamma \sum_{s_A'\in \mathcal{S}_A}P_A(s_A' | \phi^{Q^{\ast}}(s), \boldsymbol{a}) V_A^{\boldsymbol{\pi}_A^{\ast}}(s_A') \\
    &\!=\!\!\!\! \sum_{g\in G(s)} \!\!\!w(g) \!\left(\!R(g, \boldsymbol{a}) \!+ \gamma \!\!\!\sum_{s_A'\in \mathcal{S}_A}\sum_{s'\in G_A(s_A')} \!\!\!\!\!\! P(s' | g, \boldsymbol{a}) V_A^{\boldsymbol{\pi}_A^{\ast}}(s_A')\!\!\right)\!\!\! \\
    &\!=\!\!\! \sum_{g\in G(s)} \!\!w(g) \!\left(\!R(g, \boldsymbol{a}) + \gamma \sum_{s'\in \mathcal{S}}P(s' | g, \boldsymbol{a}) V_A^{\boldsymbol{\pi}_A^{\ast}}(\phi^{Q^{\ast}}(s'))\!\right).
\end{align*}}
Hence, the difference between $\sum_{g\in G(s)}w(g) Q^{\ast}(g, \boldsymbol{a})$ and $Q_A^{\boldsymbol{\pi}_A^{\ast}}(\phi^{Q^{\ast}}(s), \boldsymbol{a})$ can be bounded as follows:
{\small
\begin{align}
    &\sum_{g\in G(s)}w(g) Q^{\ast}(g, \boldsymbol{a}) - Q_A^{\boldsymbol{\pi}_A^{\ast}}(\phi^{Q^{\ast}}(s), \boldsymbol{a}) \nonumber\\
    &= \gamma\sum_{g\in G(s)} w(g) \sum_{s'\in \mathcal{S}}P(s' | g, \boldsymbol{a}) \left(V^{\ast}(s') - V_A^{\boldsymbol{\pi}_A^{\ast}}(\phi^{Q^{\ast}}(s'))\right)\! \nonumber\\
    &= \gamma \sum_{g\in G(s)}w(g)\sum_{s'\in \mathcal{S}} P(s' | g, \boldsymbol{a}) \nonumber\\
    &\phantom{=} \cdot \left(\max_{p\in \Delta(A_1)}\min_{a_2'\in \mathcal{A}_2}\sum_{a_1'\in \mathcal{A}_1}p(a_1')Q^{\ast}(s', \boldsymbol{a}') \right. \nonumber\\
    &\phantom{=} \left. - \max_{p\in \Delta(A_1)}\min_{a_2'\in \mathcal{A}_2}\sum_{a_1'\in \mathcal{A}_1}p(a_1')Q_A^{\boldsymbol{\pi}_A^{\ast}}(\phi^{Q^{\ast}}(s'), \boldsymbol{a}')\right) \nonumber\\
    &\leq \gamma \max_{(s', \boldsymbol{a}')\in \mathcal{S}\times \mathcal{A}}\left(Q^{\ast}(s', \boldsymbol{a}') - Q_A^{\boldsymbol{\pi}_A^{\ast}}(\phi^{Q^{\ast}}(s'), \boldsymbol{a}')\right),
    \label{eq:upper_bound_on_q_diff}
\end{align}}
where the second equality follows from the fact that $V^{\ast}$ and $V_A^{\boldsymbol{\pi}_A^{\ast}}$ are the minimax values in the ground and abstract TZMG, respectively.

On the other hand, under Assumption \ref{asmp:minimax_Q_abstraction}, we have the following lower bound on the state-action value difference:
{\small
\begin{align}
    &\sum_{g\in G(s)}w(g) Q^{\ast}(g, \boldsymbol{a}) - Q_A^{\boldsymbol{\pi}_A^{\ast}}(\phi^{Q^{\ast}}(s), \boldsymbol{a}) \nonumber\\
    &\geq \min_{g\in G(s)} Q^{\ast}(g, \boldsymbol{a}) - Q_A^{\boldsymbol{\pi}_A^{\ast}}(\phi^{Q^{\ast}}(s), \boldsymbol{a}) \nonumber\\
    &\geq -\epsilon + Q^{\ast}(s, \boldsymbol{a}) - Q_A^{\boldsymbol{\pi}_A^{\ast}}(\phi^{Q^{\ast}}(s), \boldsymbol{a}).
    \label{eq:lower_bound_on_q_diff}
\end{align}}

By combining \eqref{eq:upper_bound_on_q_diff} and \eqref{eq:lower_bound_on_q_diff}, and then taking the maximum value of both sides, we obtain:
{\small
\begin{align*}
    &\max_{(s, \boldsymbol{a})\in \mathcal{S}\times \mathcal{A}}\left( Q^{\ast}(s, \boldsymbol{a}) - Q_A^{\boldsymbol{\pi}_A^{\ast}}(\phi^{Q^{\ast}}(s), \boldsymbol{a})\right) \\
    &\leq \epsilon + \gamma \max_{(s, \boldsymbol{a})\in \mathcal{S}\times \mathcal{A}}\left(Q^{\ast}(s, \boldsymbol{a}) - Q_A^{\boldsymbol{\pi}_A^{\ast}}(\phi^{Q^{\ast}}(s), \boldsymbol{a})\right).
\end{align*}}
Rearranging this inequality, we have for any $s\in \mathcal{S}$ and $\boldsymbol{a}\in\mathcal{A}$:
{\small
\begin{align*}
    &Q^{\pi^{\ast}}(s, \boldsymbol{a}) - Q_A^{\boldsymbol{\pi}_A^{\ast}}(\phi^{Q^{\ast}}(s), \boldsymbol{a}) \\
    &\!\leq \!\max_{(s', \boldsymbol{a}')\in \mathcal{S}\times \mathcal{A}}\left(Q^{\pi^{\ast}}(s', \boldsymbol{a}') - Q_A^{\boldsymbol{\pi}_A^{\ast}}(\phi^{Q^{\ast}}(s'), \boldsymbol{a}')\right) \!\leq\! \frac{\epsilon}{1-\gamma}.\!
\end{align*}}
By a similar procedure, we can show that:
{\small
\begin{align*}
    &Q^{\ast}(s, \boldsymbol{a}) - Q_A^{\boldsymbol{\pi}_A^{\ast}}(\phi^{Q^{\ast}}(s), \boldsymbol{a}) \\
    &\!\geq\! \min_{(s', \boldsymbol{a}')\in \mathcal{S}\times \mathcal{A}}\left(Q^{\ast}(s', \boldsymbol{a}') \!-\! Q_A^{\boldsymbol{\pi}_A^{\ast}}(\phi^{Q^{\ast}}(s'), \boldsymbol{a}')\right) \!\geq -\frac{\epsilon}{1-\gamma}.\!
\end{align*}}
In summary, we have for any $s\in \mathcal{S}$ and $\boldsymbol{a}\in \mathcal{A}$:
{\small
\begin{align*}
    \left|Q^{\ast}(s, \boldsymbol{a}) - Q_A^{\boldsymbol{\pi}_A^{\ast}}(\phi^{Q^{\ast}}(s), \boldsymbol{a})\right| \leq \frac{\epsilon}{1-\gamma}. \qed
\end{align*}}

\paragraph{Proof of Lemma \ref{lem:diff_between_best_response_q_and_minimax_q}.}
By the definition of the state value function $V_A^{\boldsymbol{\pi}_A^{\ast}}$ in the abstract TZMG $\mathcal{M}_A$, we have for any $s\in\mathcal{S}$:
{\small
\begin{align*}
    V_A^{\boldsymbol{\pi}_A^{\ast}}(\phi^{Q^{\ast}}(s)) \!=\!\! \max_{p\in \Delta(A_1)}\min_{a_2\in \mathcal{A}_2} \!\!\sum_{a_1\in \mathcal{A}_1} \!p(a_1) Q_A^{\boldsymbol{\pi}_A^{\ast}}(\phi^{Q^{\ast}}(s), \boldsymbol{a}).
\end{align*}}
Applying Lemma \ref{lem:diff_between_q_and_q_abs_wrt_minimax_value} to this equation, we obtain the following upper bound on $V_A^{\boldsymbol{\pi}_A^{\ast}}(\phi^{Q^{\ast}}(s))$:
{\small
\begin{align*}
    V_A^{\boldsymbol{\pi}_A^{\ast}}(\phi^{Q^{\ast}}(s)) & \leq \!\! \max_{p\in \Delta(A_1)}\min_{a_2\in \mathcal{A}_2} \!\sum_{a_1\in \mathcal{A}_1} p(a_1) Q^{\ast}(s, \boldsymbol{a}) \!+\! \frac{\epsilon}{1-\gamma}\!\! \\
    & = V^{\ast}(s) + \frac{\epsilon}{1-\gamma}.
\end{align*}}
Similarly, we can derive the lower bound on $V_A^{\boldsymbol{\pi}_A^{\ast}}(\phi^{Q^{\ast}}(s))$:
{\small
\begin{align*}
    V_A^{\boldsymbol{\pi}_A^{\ast}}(\phi^{Q^{\ast}}(s)) & \geq \!\!\max_{p\in \Delta(A_1)}\min_{a_2\in \mathcal{A}_2} \!\sum_{a_1\in \mathcal{A}_1} p(a_1) Q^{\ast}(s, \boldsymbol{a}) \!-\! \frac{\epsilon}{1-\gamma} \!\! \\
    &= V^{\ast}(s) - \frac{\epsilon}{1-\gamma}.
\end{align*}}
Hence, we have for any $s\in \mathcal{S}$:
{\small
\begin{align*}
    \left|V^{\ast}(s) - V_A^{\boldsymbol{\pi}_A^{\ast}}(\phi^{Q^{\ast}}(s))\right| \leq \frac{\epsilon}{1-\gamma}.
\end{align*}}
By using this inequality, we obtain for any $s\in \mathcal{S}$ and $\boldsymbol{a}\in \mathcal{A}$,
{\small
\begin{align}
    &Q^{\pi_1^{\dagger}, \pi_{GA, 2}^{\ast}}(s, \boldsymbol{a}) - Q^{\ast}(s, \boldsymbol{a}) \nonumber \\
    &= \gamma \sum_{s'\in \mathcal{S}}P(s' | s, \boldsymbol{a}) \left( V^{\pi_1^{\dagger}, \pi_{GA, 2}^{\ast}}(s') - V^{\ast}(s')\right) \nonumber\\
    &= \gamma \sum_{s'\in \mathcal{S}}P(s' | s, \boldsymbol{a}) \left(V_A^{\boldsymbol{\pi}_A^{\ast}}(\phi^{Q^{\ast}}(s')) - V^{\ast}(s')\right) \nonumber\\
    &\phantom{=} + \gamma \sum_{s'\in \mathcal{S}}P(s' | s, \boldsymbol{a}) \left(V^{\pi_1^{\dagger}, \pi_{GA, 2}^{\ast}}(s') - V_A^{\boldsymbol{\pi}_A^{\ast}}(\phi^{Q^{\ast}}(s'))\right) \nonumber\\
    &\leq \frac{\gamma \epsilon}{1-\gamma} \nonumber\\
    &\phantom{=} + \gamma \sum_{s'\in \mathcal{S}}P(s' | s, \boldsymbol{a}) \left(V^{\pi_1^{\dagger}, \pi_{GA, 2}^{\ast}}(s') - V_A^{\boldsymbol{\pi}_A^{\ast}}(\phi^{Q^{\ast}}(s'))\right),
    \label{eq:upper_bound_on_diff_between_best_response_q_and_minimax_q}
\end{align}}%
\noindent where the first equality follows from the definition of the state-action value function.
Here, we can upper bound the term of $V^{\pi_1^{\dagger}, \pi_{GA, 2}^{\ast}}(s') - V_A^{\boldsymbol{\pi}_A^{\ast}}(\phi^{Q^{\ast}}(s'))$ as follows:
{\small
\begin{align}
    & V^{\pi_1^{\dagger}, \pi_{GA, 2}^{\ast}}(s') - V_A^{\boldsymbol{\pi}_A^{\ast}}(\phi^{Q^{\ast}}(s')) \nonumber\\
    &= \max_{a_1'\in \mathcal{A}_1}\sum_{a_2'\in \mathcal{A}_2} \pi_{GA, 2}^{\ast}(a_2'|s')Q^{\pi_1^{\dagger}, \pi_{GA, 2}^{\ast}}(s', \boldsymbol{a}')  \nonumber\\
    &\phantom{=} - \max_{a_1'\in \mathcal{A}_1}\sum_{a_2'\in \mathcal{A}_2} \pi_{GA,2}^{\ast}(a_2'|s')Q_A^{\boldsymbol{\pi}_A^{\ast}}(\phi^{Q^{\ast}}(s'), \boldsymbol{a}') \nonumber\\
    &\leq \max_{\boldsymbol{a}'\in \mathcal{A}}\left( Q^{\pi_1^{\dagger}, \pi_{GA, 2}^{\ast}}(s', \boldsymbol{a}') - Q_A^{\boldsymbol{\pi}_A^{\ast}}(\phi^{Q^{\ast}}(s'), \boldsymbol{a}')\right).
    \label{eq:upper_bound_on_diff_between_best_response_v_and_abstract_minimax_v}
\end{align}}
By combining \eqref{eq:upper_bound_on_diff_between_best_response_q_and_minimax_q}, \eqref{eq:upper_bound_on_diff_between_best_response_v_and_abstract_minimax_v}, and Lemma \ref{lem:diff_between_q_and_q_abs_wrt_minimax_value}, we have for any $s\in \mathcal{S}$ and $\boldsymbol{a}\in \mathcal{A}$:
{\small
\begin{align*}
    &Q^{\pi_1^{\dagger}, \pi_{GA, 2}^{\ast}}(s, \boldsymbol{a}) - Q^{\ast}(s, \boldsymbol{a}) \leq \frac{2\gamma \epsilon}{1-\gamma} \\
    &\phantom{=} + \gamma \max_{(s', \boldsymbol{a}')\in \mathcal{S}\times \mathcal{A}}\left( Q^{\pi_1^{\dagger}, \pi_{GA, 2}^{\ast}}(s', \boldsymbol{a}') - Q^{\ast}(s', \boldsymbol{a}')\right).
\end{align*}}
Taking the maximum value of both sides:
{\small
\begin{align*}
    &\max_{(s, \boldsymbol{a})\in \mathcal{S}\times \mathcal{A}} \left(Q^{\pi_1^{\dagger}, \pi_{GA, 2}^{\ast}}(s, \boldsymbol{a}) - Q^{\pi^{\ast}}(s, \boldsymbol{a})\right) \\
    &\leq \frac{2\gamma \epsilon}{1-\gamma} + \gamma \max_{(s, \boldsymbol{a})\in \mathcal{S}\times \mathcal{A}} \left(Q^{\pi_1^{\dagger}, \pi_{GA, 2}^{\ast}}(s, \boldsymbol{a}) - Q^{\pi^{\ast}}(s, \boldsymbol{a})\right).
\end{align*}}
Rearranging this inequality, we have for any $s\in \mathcal{S}$ and $\boldsymbol{a}\in \mathcal{A}$:
{\small
\begin{align}
    &Q^{\pi_1^{\dagger}, \pi_{GA, 2}^{\ast}}(s, \boldsymbol{a}) - Q^{\ast}(s, \boldsymbol{a}) \nonumber\\
    &\!\leq\! \max_{(s', \boldsymbol{a}')\in \mathcal{S}\times \mathcal{A}} \!\left(\!Q^{\pi_1^{\dagger}, \pi_{GA, 2}^{\ast}}(s', \boldsymbol{a}') - Q^{\ast}(s', \boldsymbol{a}')\!\right) \!\leq\! \frac{2\epsilon}{(1-\gamma)^2}.
    \label{eq:lower_bound_on_diff_between_minimax_q_and_best_response_q}
\end{align}}

On the other hand, from the property of minimax state-action value in \eqref{eq:shapley_theorem}, we have: 
{\small
\begin{align}
    Q^{\ast}(s, \boldsymbol{a}) - Q^{\pi_1^{\dagger}, \pi_{GA, 2}^{\ast}}(s, \boldsymbol{a}) \leq 0.
    \label{eq:upper_bound_on_diff_between_minimax_q_and_best_response_q}
\end{align}}
By combining \eqref{eq:lower_bound_on_diff_between_minimax_q_and_best_response_q} with \eqref{eq:upper_bound_on_diff_between_minimax_q_and_best_response_q}, we get for any $s\in \mathcal{S}$ and $\boldsymbol{a}\in \mathcal{A}$:
{\small
\begin{align*}
    \left|Q^{\pi_1^{\dagger}, \pi_{GA, 2}^{\ast}}(s, \boldsymbol{a}) - Q^{\ast}(s, \boldsymbol{a})\right| \leq \frac{2\epsilon}{(1-\gamma)^2}.\qed
\end{align*}}

\section{Experiments}\label{sec:experiments}
This section demonstrates our state abstraction developed so far in Markov Soccer~\cite{littman1994markov,abe2021off}. We here focus only on the minimax values since the theoretical results on the other criteria in Section~\ref{sec:discussion} rely on the minimax values. 
\subsection{Markov Soccer}
\begin{figure}[t]
    \centering
    \includegraphics[width=0.5\linewidth]{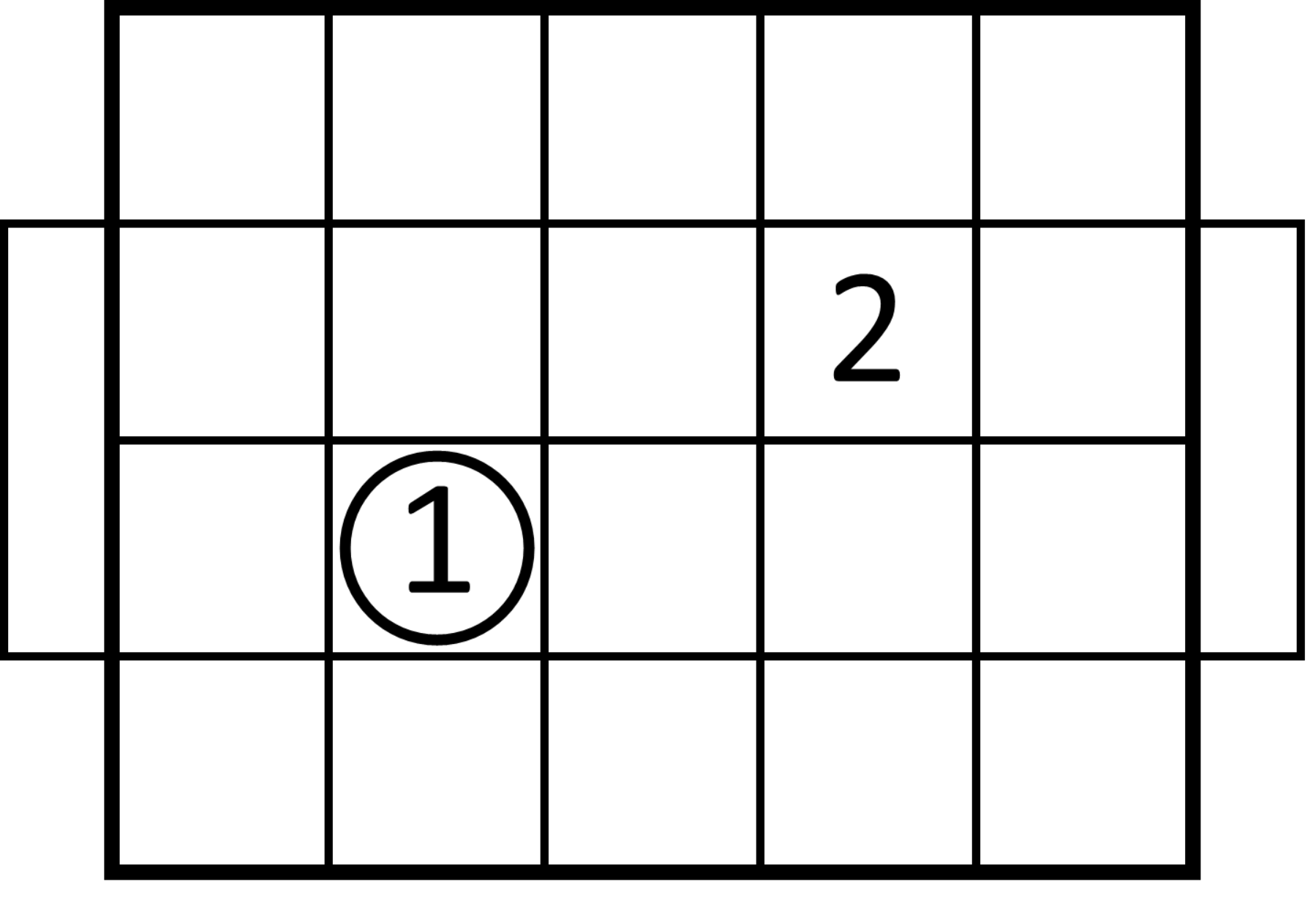}
    \caption{An initial state of the Markov soccer game in which player 1 has the ball.}
    \label{fig:markov_soccer_initial_state}
\end{figure}

We describe the Markov soccer experiment as 1 vs 1 game on $4\times 5$ as shown in Figure~\ref{fig:markov_soccer_initial_state}. Two players ``1'' and ``2'' occupy distinct squares of the grid, respectively and the circled player, ``1'' here, has a ``ball,'' which specifies the states of the game. Figure~\ref{fig:markov_soccer_initial_state} shows the initial positions of the players. Which player has the ball at the initial turn is determined at uniformly random. In each turn, each player can move to one of the neighboring cells or stand at the place, i.e., their set of actions includes ``Up'', ``Left'', ``Down'', ``Right'', and ``Stand.'' 

After both select their actions, these two moves are executed in random order. When a player tries to move to the cell occupied by the other player, the ball’s possession goes to the stationary player, and the positions of both players remain unchanged. 
Also, if a player's choice lets him or her be out of the pitch, the position of the player remains unchanged.
When a player keeping the ball steps into his or her goal (the right side for player~1 and the left side for player~2), the game is over. At the same time, that player scores 1 point and the opponent scores -1 point.
The positions of the players and the ball's possession are initialized as shown in Figure~\ref{fig:markov_soccer_initial_state}.

\subsection{Training and Evaluating}
In Markov soccer, we build state abstraction, by first solving the game, then greedily aggregating ground states into abstract states that satisfy the $Q^{\ast}$ criterion. 
We calculate the number of states in the abstract Markov game $|\mathcal{S}_A|$, varying $\epsilon$ from $0.0$ to $2.0$ in increments of $0.1$.  
Then, for each $\epsilon$, we compute the equilibrium policies ${\boldsymbol \pi}^\ast$ and ${\boldsymbol \pi}_A^\ast$ for the ground and the abstract Markov soccer games, respectively, via the minimax Q-learning. 
We here assume that the total number of learning iterations $T$ was 1,000,000, the discount factor $\gamma$ was 0.9, and the learning rate $\alpha_t$ was set to $10^{-\frac{2}{T}t}$ for learning iterations $t \geq 0$.
We further evaluate the duality gaps for those equilibrium policies. However, it is demanding to calculate the true one, so we use the approximation obtained from Q-learning. 

\subsection{Results}

\begin{figure}[t]
    \centering
    \includegraphics[width=0.7\linewidth]{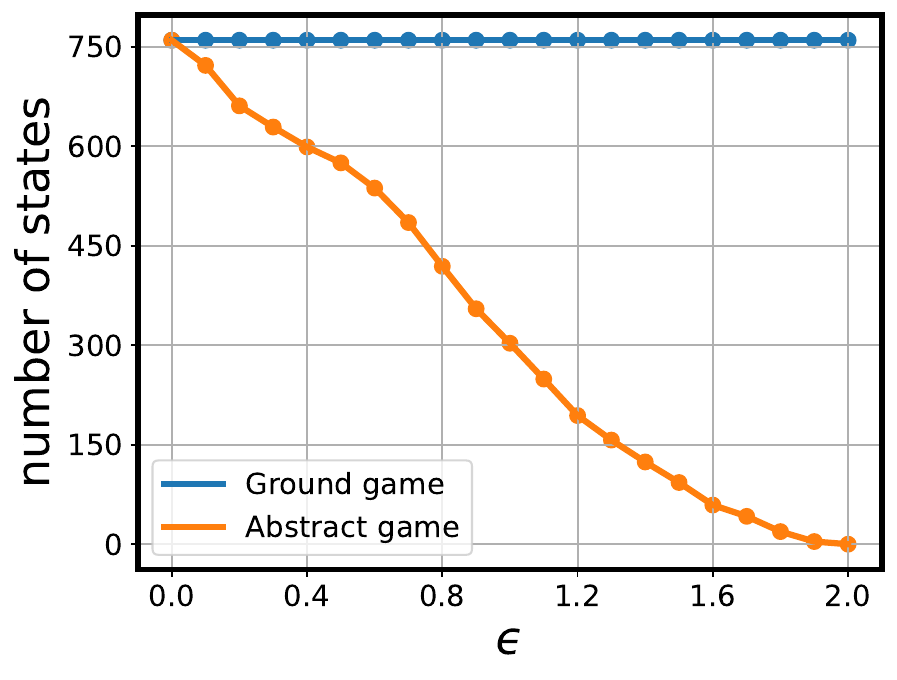}
    \caption{Number of states in the abstract Markov soccer games with respect to $\epsilon$.}
    \label{fig:state_size_plot}
\end{figure}

We first compute the number of states in the abstract Markov soccer game for different of $\epsilon$ in Figure~\ref{fig:state_size_plot}. The x- and y-axes represent $\epsilon$ and the number of states $|\mathcal{S}_A|$, respectively. We observed that as $\epsilon$ increases, the number of states decreases almost linearly. Note that the number of states of the ground Markov soccer is $760$, as illustrated at $\epsilon=0.0$. We observe the value of $\epsilon$ up to two since the difference between the state-action value functions is bound up to $2$ from the definition of the game. The number of states of the abstract Markov soccer is $1$ at $\epsilon=2.0$.

% '<insert vs. duality gap' >

\begin{figure}[t]
    \centering
    \includegraphics[width=0.7\linewidth]{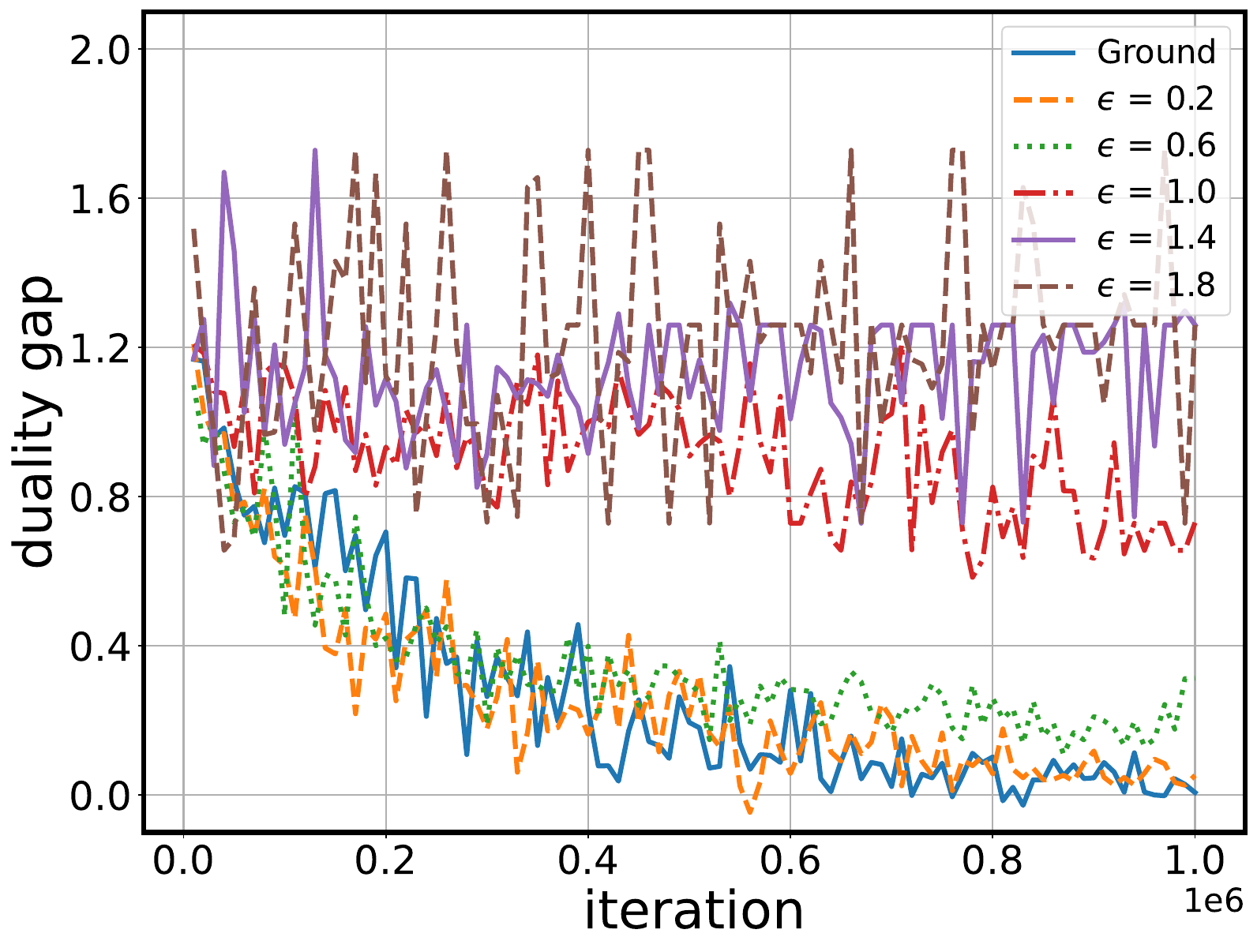}
    \caption{Duality gap at each iteration in minimax Q-learning. Note that the policies are trained in the abstract game, and their duality gap values are computed in the ground game.}
    \label{fig:exploitability_custom_1000000}
\end{figure}

Figure \ref{fig:exploitability_custom_1000000} illustrates the duality gap in the number of learning iterations where x- and y-axes represent learning iterations and the gap, respectively, varying $\epsilon$. 
We label ``Ground'' the gap for the ground Markov soccer game ($\epsilon=0.0$) and draw the trajectories varying $\epsilon\in\{0.2, 0.6, 1.0, 1.4, 1.8\}$. We observe that the minimax Q-learning approximately solves the ground game, i.e., the gap converges to zero. If $\epsilon$ is sufficiently small, i.e., less than $0.6$, the duality gap in the abstract Markov soccer game approximates the ground one. Otherwise, the gaps become significantly worse. 
In these cases, agents often repeat previous actions without progress. 
% For example, one agent may choose to remain stationary, while the other moves down according to the optimal policy of the abstracted game. If this agent is positioned right above a wall or a boundary in the ground game, it is unable to proceed, leading both agents to continue in their respective positions as part of a mutual best response. 
Such deadlocks increase the duality gap, as agents can more easily identify best responses by fully exploring the ground game’s strategy space. 
% We can argue that our approach works well for some appropriate $\epsilon$. 

\section{Extentions}\label{sec:discussion}
We have extended the approximate function that preserves near-optimal behavior by aggregating states on similar optimal Q-values to two-player zero-sum Markov games. This section further extends three other types of criteria: \textit{Model Similarity}; \textit{Boltzmann Distribution Similarity}; \textit{Multinomial Distribution similarity}. 
%
%similarity of rewards and transitions (\textit{Model Similarity}); similarity of a Boltzmann distribution over optimal Q-values (\textit{Boltzmann Distribution Similarity}); and similarity of a multinomial distribution over optimal Q-values (\textit{Multinomial Distribution similarity}). 
We then % successfully 
derive the bounds of the gap function for each criterion.
Since the theoretical results on these three criteria rely on the one on the minimax Q-values, the proofs are placed in Appendix~\ref{sec:model_similarity_proof}-\ref{sec:multinomial_similarity_proof}.

% \subsection{Model Similarity}
Let us now consider Model Similarity where states are aggregated together when their rewards and transitions are within $\epsilon$~\cite{li2006towards,abel2016near}. Specifically, when a aggregation function $\phi$ aggregates states $s_1, s_2 \in \mathcal{S}$, the difference between the probability functions $P(s_1,\boldsymbol{a})$ and $P(s_2,\boldsymbol{a})$ is bounded within $\epsilon\in[0,\infty)$. As well, the difference between the reward functions $R(s_1,\boldsymbol{a})$ and $R(s_2,\boldsymbol{a})$ is bounded within the same error amount. 

\begin{assumption}
\label{asmp:error_wrt_model_similarity}
The aggregation function $\phi^{\mathrm{model}}$ satisfies the following property for some non-negative constant~$\epsilon\geq~0$: $\phi^{\mathrm{model}}(s_1) = \phi^{\mathrm{model}}(s_2) \Rightarrow$%
{\small
\begin{align*}
    % &\phi^{\mathrm{model}}(s_1) = \phi^{\mathrm{model}}(s_2) \Rightarrow \\ 
    &\max_{\boldsymbol{a}\in \mathcal{A}} \left|R(s_1, \boldsymbol{a}) - R(s_2, \boldsymbol{a})\right| \leq \epsilon,\ \mathrm{and} \\
    &\phantom{\Rightarrow}\phantom{and } \max_{s_A'\in \mathcal{S}_A}\left|\sum_{s'\in G_A(s_A')} \left(P(s' | s_1, \boldsymbol{a}) - P(s' | s_2, \boldsymbol{a})\right)\right| \leq \epsilon.
\end{align*}}
\end{assumption}
We derive the bound of the duality gap for the Model Similarity criteria as follows. 
\begin{theorem}\label{thm:error_wrt_model_similarity}
When the ground states are aggregated by the aggregation function $\phi^{\mathrm{model}}$ satisfying Assumption \ref{asmp:error_wrt_model_similarity} with $\epsilon \geq 0$, then $\boldsymbol{\pi}_{GA}^{\ast}$ satisfies:
{\small
\begin{align*}
    \mathrm{GAP}\left(\boldsymbol{\pi}_{GA}^{\ast}\right) \leq \frac{4(1 + \gamma \left(|\mathcal{S}|- 1\right))\epsilon}{(1-\gamma)^3}.
\end{align*}}
\end{theorem}

% \subsection{Boltzmann Distribution Similarity}
Let us next examine Boltzmann Distribution Similarity from the classic textbook~\cite{sutton:1998}, which balances exploration and exploitation and allows for aggregating states when the ratios of Q-values are similar, but the magnitudes are different. 
\begin{assumption}
\label{asmp:error_wrt_boltzmann}
The aggregation function $\phi^{\mathrm{bolt}}$ satisfies the following property for some non-negative constants $\epsilon\geq 0$ and $k\geq 0$: $\phi^{\mathrm{bolt}}(s_1) = \phi^{\mathrm{bolt}}(s_2)\Rightarrow$
{\small
\begin{align*}
    % &\phi^{\mathrm{bolt}}(s_1) = \phi^{\mathrm{bolt}}(s_2) \nonumber\\
    & \max_{\boldsymbol{a}\in \mathcal{A}}\left|\frac{e^{Q^{\ast}(s_1, \boldsymbol{a})}}{\sum_{\boldsymbol{b}\in \mathcal{A}}e^{Q^{\ast}(s_1, \boldsymbol{b})}} - \frac{e^{Q^{\ast}(s_2, \boldsymbol{a})}}{\sum_{\boldsymbol{b}\in \mathcal{A}}e^{Q^{\ast}(s_2, \boldsymbol{b})}}\right| \leq \epsilon,\ \mathrm{and\ } \nonumber\\
    &\phantom{\Rightarrow} \left|\sum_{\boldsymbol{b}\in \mathcal{A}}e^{Q^{\ast}(s_1, \boldsymbol{b})} - \sum_{\boldsymbol{b}\in \mathcal{A}}e^{Q^{\ast}(s_2, \boldsymbol{b})}\right| \leq k\epsilon.
\end{align*}}
\end{assumption}
\begin{theorem}\label{thm:error_wrt_boltzmann_similarity}
When the ground states are aggregated by the aggregation function $\phi^{\mathrm{bolt}}$ satisfying Assumption \ref{asmp:error_wrt_boltzmann} with $\epsilon \geq 0$ and $k\geq 0$, then $\boldsymbol{\pi}_{GA}^{\ast}$ satisfies:
{\small
\begin{align*}
    \mathrm{GAP}\left(\boldsymbol{\pi}_{GA}^{\ast}\right) \leq \frac{12e^{\frac{2}{1-\gamma}}\left(|\mathcal{A}_1||\mathcal{A}_2| + k\frac{e^{\frac{1}{1-\gamma}}}{|\mathcal{A}_1||\mathcal{A}_2|}\right) \epsilon}{(1-\gamma)^3}.
\end{align*}}
\end{theorem}% 

% \subsection{Multinomial Distribution Similarity}
We finally consider Multinomial Distribution similarity, which is a bit simpler and has close properties to Boltzmann Distribution Similarity. 
\begin{assumption}
\label{asmp:error_wrt_multinomial}
The aggregation function $\phi^{\mathrm{mult}}$ satisfies the following property for some non-negative constants $\epsilon\geq 0$ and $k\geq 0$: $\phi^{\mathrm{mult}}(s_1) = \phi^{\mathrm{mult}}(s_2)\Rightarrow$
{\small
\begin{align*}
    % &\phi^{\mathrm{mult}}(s_1) = \phi^{\mathrm{mult}}(s_2) \nonumber\\
    &\max_{\boldsymbol{a}\in \mathcal{A}} \left|\frac{Q^{\ast}(s_1, \boldsymbol{a})}{\sum_{\boldsymbol{b}\in \mathcal{A}}Q^{\ast}(s_1, \boldsymbol{b})} - \frac{Q^{\ast}(s_2, \boldsymbol{a})}{\sum_{\boldsymbol{b}\in \mathcal{A}}Q^{\ast}(s_2, \boldsymbol{b})}\right| \leq \epsilon,\ \mathrm{and\ } \nonumber\\
    &\phantom{\Rightarrow} \left|\sum_{\boldsymbol{b}\in \mathcal{A}}Q^{\ast}(s_1, \boldsymbol{b}) - \sum_{\boldsymbol{b}\in \mathcal{A}}Q^{\ast}(s_2, \boldsymbol{b})\right| \leq k\epsilon.
\end{align*}}
\end{assumption}
\begin{theorem}
\label{thm:error_wrt_multinomial}
Suppose that the ground states are aggregated by the aggregation function $\phi^{\mathrm{mult}}$ satisfying Assumption \ref{asmp:error_wrt_multinomial} with $\epsilon \geq 0$ and $k\geq 0$.
If there exists some positive constant $\delta > 0$ such that $\left|\sum_{\boldsymbol{b}\in \mathcal{A}}Q^{\ast}(s, \boldsymbol{b})\right| \geq \delta$ for any states $s\in \mathcal{S}$, 
{\small
\begin{align*}
    \mathrm{GAP}\left(\boldsymbol{\pi}_{GA}^{\ast}\right) \leq \frac{12\left(|\mathcal{A}_1||\mathcal{A}_2| + \frac{k}{\delta}\right)\epsilon}{(1-\gamma)^4}.
\end{align*}}
\end{theorem}

\section{Conclusion}
This paper investigates approximate state abstraction, which was originally developed for single-agent MDPs, and extends it for TZMGs, 
% where two players have conflicts of interest and 
which potentially have many real-world applications. 
Future works include conducting experiments on % the other criteria and 
games with larger state spaces, such as ``The Chasing Game on Gridworld~\cite{wang:icml:2018}'' and ``Snake Games~\cite{guibas:iclr:2022}.''

\section*{Acknowledgments}
We would like to thank Yuki Shimano for useful discussions. Atsushi Iwasaki is supported by JSPS KAKENHI Grant Numbers 21H04890 and 23K17547.

\bibliography{aaai25}

\begin{thebibliography}{34}
\providecommand{\natexlab}[1]{#1}

\bibitem[{Abe and Kaneko(2021)}]{abe2021off}
Abe, K.; and Kaneko, Y. 2021.
\newblock Off-Policy Exploitability-Evaluation in Two-Player Zero-Sum Markov Games.
\newblock In \emph{AAMAS}, 78--87.

\bibitem[{Abel, Hershkowitz, and Littman(2016)}]{abel2016near}
Abel, D.; Hershkowitz, D.~E.; and Littman, M.~L. 2016.
\newblock Near optimal behavior via approximate state abstraction.
\newblock In \emph{ICML}, 2915--2923.

\bibitem[{Brown and Sandholm(2018)}]{brown:science:2018}
Brown, N.; and Sandholm, T. 2018.
\newblock Superhuman AI for heads-up no-limit poker: Libratus beats top professionals.
\newblock \emph{Science}, 359(6374): 418--424.

\bibitem[{Burch, Johanson, and Bowling(2014)}]{burch2014solving}
Burch, N.; Johanson, M.; and Bowling, M. 2014.
\newblock Solving imperfect information games using decomposition.
\newblock In \emph{AAAI}, 602--608.

\bibitem[{Castro(2020)}]{castro2020scalable}
Castro, P.~S. 2020.
\newblock Scalable Methods for Computing State Similarity in Deterministic Markov Decision Processes.
\newblock In \emph{AAAI}, 10069--10076.

\bibitem[{Dadvar, Nayyar, and Srivastava(2023)}]{dadvar2023conditional}
Dadvar, M.; Nayyar, R.~K.; and Srivastava, S. 2023.
\newblock Conditional abstraction trees for sample-efficient reinforcement learning.
\newblock In \emph{UAI}, 485--495.

\bibitem[{Dietterich(1998)}]{dietterich1998maxq}
Dietterich, T. 1998.
\newblock The MAXQ Method for Hierarchical Reinforcement Learning.
\newblock In \emph{ICML}, 118--126.

\bibitem[{Dietterich(1999)}]{dietterich1999state}
Dietterich, T. 1999.
\newblock State abstraction in MAXQ hierarchical reinforcement learning.
\newblock In \emph{NeurIPS}, 994--1000.

\bibitem[{Ferns, Panangaden, and Precup(2004)}]{ferns2004metrics}
Ferns, N.; Panangaden, P.; and Precup, D. 2004.
\newblock Metrics for finite Markov decision processes.
\newblock In \emph{UAI}, 162--169.

\bibitem[{Ganzfried and Sandholm(2013)}]{ganzfried2013action}
Ganzfried, S.; and Sandholm, T. 2013.
\newblock Action translation in extensive-form games with large action spaces: axioms, paradoxes, and the pseudo-harmonic mapping.
\newblock In \emph{IJCAI}, 120--128.

\bibitem[{Gilpin(2006)}]{gilpin2006competitive}
Gilpin, A. 2006.
\newblock A Competitive Texas Hold'em Poker Player via Automated Abstraction and Real-Time Equilibrium Computation.
\newblock In \emph{AAAI}, 1007--1013.

\bibitem[{Gilpin and Sandholm(2006)}]{gilpin2006finding}
Gilpin, A.; and Sandholm, T. 2006.
\newblock Finding equilibria in large sequential games of imperfect information.
\newblock In \emph{EC}, 160--169.

\bibitem[{Gilpin and Sandholm(2007)}]{gilpin2007better}
Gilpin, A.; and Sandholm, T. 2007.
\newblock Better automated abstraction techniques for imperfect information games, with application to Texas Hold'em poker.
\newblock In \emph{AAMAS}, 1--8.

\bibitem[{Gilpin, Sandholm, and S\o{}rensen(2007)}]{gilpin2007potential}
Gilpin, A.; Sandholm, T.; and S\o{}rensen, T.~B. 2007.
\newblock Potential-aware automated abstraction of sequential games, and holistic equilibrium analysis of Texas Hold'em poker.
\newblock In \emph{AAAI}, 50--57.

\bibitem[{Givan, Dean, and Greig(2003)}]{givan2003equivalence}
Givan, R.; Dean, T.; and Greig, M. 2003.
\newblock Equivalence notions and model minimization in Markov decision processes.
\newblock \emph{Artificial intelligence}, 147(1–2): 163--223.

\bibitem[{Guibas et~al.(2022)Guibas, Mardani, Li, Tao, Anandkumar, and Catanzaro}]{guibas:iclr:2022}
Guibas, J.; Mardani, M.; Li, Z.; Tao, A.; Anandkumar, A.; and Catanzaro, B. 2022.
\newblock Efficient Token Mixing for Transformers via Adaptive Fourier Neural Operators.
\newblock In \emph{ICLR}.

\bibitem[{Johanson et~al.(2013)Johanson, Burch, Valenzano, and Bowling}]{johanson2013evaluating}
Johanson, M.; Burch, N.; Valenzano, R.; and Bowling, M. 2013.
\newblock Evaluating state-space abstractions in extensive-form games.
\newblock In \emph{AAMAS}, 271--278.

\bibitem[{Jong and Stone(2005)}]{jong:ijcai:2005}
Jong, N.~K.; and Stone, P. 2005.
\newblock State abstraction discovery from irrelevant state variables.
\newblock In \emph{IJCAI}, 752--–757.

\bibitem[{Jonsson and Barto(2000)}]{jonsson2000automated}
Jonsson, A.; and Barto, A.~G. 2000.
\newblock Automated state abstraction for options using the U-Tree algorithm.
\newblock In \emph{NeurIPS}, 1010--1016.

\bibitem[{Kroer and Sandholm(2018)}]{kroer2018unified}
Kroer, C.; and Sandholm, T. 2018.
\newblock A unified framework for extensive-form game abstraction with bounds.
\newblock In \emph{NeurIPS}, 613--624.

\bibitem[{Li, Walsh, and Littman(2006)}]{li2006towards}
Li, L.; Walsh, T.~J.; and Littman, M.~L. 2006.
\newblock Towards a Unified Theory of State Abstraction for MDPs.
\newblock In \emph{ISAIM}.

\bibitem[{Littman(1994)}]{littman1994markov}
Littman, M.~L. 1994.
\newblock Markov games as a framework for multi-agent reinforcement learning.
\newblock In \emph{ICML}, 157--163.

\bibitem[{Perolat et~al.(2022)Perolat, Vylder, Hennes, Tarassov, Strub, de~Boer, Muller, Connor, Burch, Anthony, McAleer, Elie, Cen, Wang, Gruslys, Malysheva, Khan, Ozair, Timbers, Pohlen, Eccles, Rowland, Lanctot, Lespiau, Piot, Omidshafiei, Lockhart, Sifre, Beauguerlange, Munos, Silver, Singh, Hassabis, and Tuyls}]{perolat:science:2022}
Perolat, J.; Vylder, B.~D.; Hennes, D.; Tarassov, E.; Strub, F.; de~Boer, V.; Muller, P.; Connor, J.~T.; Burch, N.; Anthony, T.; McAleer, S.; Elie, R.; Cen, S.~H.; Wang, Z.; Gruslys, A.; Malysheva, A.; Khan, M.; Ozair, S.; Timbers, F.; Pohlen, T.; Eccles, T.; Rowland, M.; Lanctot, M.; Lespiau, J.-B.; Piot, B.; Omidshafiei, S.; Lockhart, E.; Sifre, L.; Beauguerlange, N.; Munos, R.; Silver, D.; Singh, S.; Hassabis, D.; and Tuyls, K. 2022.
\newblock Mastering the game of Stratego with model-free multiagent reinforcement learning.
\newblock \emph{Science}, 378(6623): 990--996.

\bibitem[{Ravindran and Barto(2003)}]{ravindran2003smdp}
Ravindran, B.; and Barto, A.~G. 2003.
\newblock SMDP homomorphisms: an algebraic approach to abstraction in semi-Markov decision processes.
\newblock In \emph{IJCAI}, 1011--1016.

\bibitem[{Ravindran and Barto(2004)}]{ravindran2004approximate}
Ravindran, B.; and Barto, A.~G. 2004.
\newblock Approximate homomorphisms: A framework for non-exact minimization in Markov decision processes.
\newblock In \emph{International Conference on Knowledge Based Computer Systems}, 19--22.

\bibitem[{Shalev-Shwartz, Shammah, and Shashua(2016)}]{shalev::2016}
Shalev-Shwartz, S.; Shammah, S.; and Shashua, A. 2016.
\newblock Safe, multi-agent, reinforcement learning for autonomous driving.
\newblock \emph{arXiv preprint arXiv:1610.03295}.

\bibitem[{Shapley(1953)}]{shapley1953stochastic}
Shapley, L.~S. 1953.
\newblock Stochastic Games.
\newblock \emph{Proceedings of the National Academy of Sciences}, 39(10): 1095--1100.

\bibitem[{Silver et~al.(2016)Silver, Huang, Maddison, Guez, Sifre, van~den Driessche, Schrittwieser, Antonoglou, Panneershelvam, Lanctot, Dieleman, Grewe, Nham, Kalchbrenner, Sutskever, Lillicrap, Leach, Kavukcuoglu, Graepel, and Hassabis}]{silver2016mastering}
Silver, D.; Huang, A.; Maddison, C.~J.; Guez, A.; Sifre, L.; van~den Driessche, G.; Schrittwieser, J.; Antonoglou, I.; Panneershelvam, V.; Lanctot, M.; Dieleman, S.; Grewe, D.; Nham, J.; Kalchbrenner, N.; Sutskever, I.; Lillicrap, T.~P.; Leach, M.; Kavukcuoglu, K.; Graepel, T.; and Hassabis, D. 2016.
\newblock Mastering the game of Go with deep neural networks and tree search.
\newblock \emph{Nature}, 529(7587): 484--489.

\bibitem[{Silver et~al.(2017)Silver, Schrittwieser, Simonyan, Antonoglou, Huang, Guez, Hubert, Baker, Lai, Bolton, Chen, Lillicrap, Hui, Sifre, van~den Driessche, Graepel, and Hassabis}]{silver2017mastering}
Silver, D.; Schrittwieser, J.; Simonyan, K.; Antonoglou, I.; Huang, A.; Guez, A.; Hubert, T.; Baker, L.; Lai, M.; Bolton, A.; Chen, Y.; Lillicrap, T.~P.; Hui, F.; Sifre, L.; van~den Driessche, G.; Graepel, T.; and Hassabis, D. 2017.
\newblock Mastering the game of Go without human knowledge.
\newblock \emph{Nature}, 550(7676): 354--359.

\bibitem[{Sutton and Barto(1998)}]{sutton:1998}
Sutton, R.~S.; and Barto, A.~G. 1998.
\newblock \emph{Reinforcement Learning: An Introduction}.
\newblock MIT Press.

\bibitem[{Szepesvári and Littman(1999)}]{szepesvari1999unified}
Szepesvári, C.; and Littman, M.~L. 1999.
\newblock A Unified Analysis of Value-Function-Based Reinforcement-Learning Algorithms.
\newblock \emph{Neural Computation}, 11(8): 2017--2060.

\bibitem[{van~der Pol et~al.(2020)van~der Pol, Kipf, Oliehoek, and Welling}]{van2020plannable}
van~der Pol, E.; Kipf, T.; Oliehoek, F.~A.; and Welling, M. 2020.
\newblock Plannable Approximations to MDP Homomorphisms: Equivariance under Actions.
\newblock In \emph{AAMAS}, 1431--1439.

\bibitem[{Wang and Klabjan(2018)}]{wang:icml:2018}
Wang, X.; and Klabjan, D. 2018.
\newblock Competitive Multi-agent Inverse Reinforcement Learning with Sub-optimal Demonstrations.
\newblock In \emph{ICML}, 5143--5151.

\bibitem[{Waugh(2013)}]{waugh2013fast}
Waugh, K. 2013.
\newblock A fast and optimal hand isomorphism algorithm.
\newblock In \emph{AAAI Workshop on Computer Poker and Imperfect Information}.

\end{thebibliography}

\newpage
\clearpage

% \begin{comment}

\appendix
\onecolumn

\section{Proofs for Theorem \ref{thm:error_wrt_minimax_value}}
\label{sec:minimax_value}

\subsection{Proof of Lemma  \ref{lem:exploitability_bound_minimax_value}}
\label{app:exploitability_bound_minimax_value}
\begin{proof}[Proof of Lemma \ref{lem:exploitability_bound_minimax_value}]
First, we introduce the following lemma:
\begin{lemma}
\label{lem:v_to_q}
Assume that the aggregation function satisfies the following condition for some non-negative constant $\delta \geq 0$:
\begin{align*}
    \forall s\in \mathcal{S}, \boldsymbol{a}\in \mathcal{A}, ~\left|Q^{\pi_1^{\dagger}, \pi_{GA, 2}^{\ast}}(s, \boldsymbol{a}) - Q_A^{\boldsymbol{\pi}_A^{\ast}}(\phi(s), \boldsymbol{a})\right| \leq \delta.
\end{align*}
Then, for any $s\in \mathcal{S}$, we have:
\begin{align*}
    V^{\pi_1^{\dagger}, \pi_{GA, 2}^{\ast}}(s) \leq \sum_{\boldsymbol{a}\in \mathcal{A}}\boldsymbol{\pi}_{GA}^{\ast}(\boldsymbol{a} | s)Q^{\pi_1^{\dagger}, \pi_{GA, 2}^{\ast}}(s, \boldsymbol{a}) + 2\delta.
\end{align*}
\end{lemma}
The proof of this lemma is shown in Appendix \ref{app:v_to_q}.
From Lemma \ref{lem:v_to_q}, we have for any $s\in \mathcal{S}$:
\begin{align*}
    V^{\pi_1^{\dagger}, \pi_{GA, 2}^{\ast}}(s) - V^{\boldsymbol{\pi}_{GA}^{\ast}}(s) &\leq 2\delta + \sum_{\boldsymbol{a}\in \mathcal{A}}\boldsymbol{\pi}_{GA}^{\ast}(\boldsymbol{a} | s)Q^{\pi_1^{\dagger}, \pi_{GA, 2}^{\ast}}(s, \boldsymbol{a})  - V^{\boldsymbol{\pi}_{GA}^{\ast}}(s) \\
    &= 2\delta + \sum_{\boldsymbol{a}\in \mathcal{A}}\boldsymbol{\pi}_{GA}^{\ast}(\boldsymbol{a} | s)\left(Q^{\pi_1^{\dagger}, \pi_{GA, 2}^{\ast}}(s, \boldsymbol{a}) - Q^{\boldsymbol{\pi}_{GA}^{\ast}}(s, \boldsymbol{a})\right) \\
    &= 2\delta + \gamma \sum_{\boldsymbol{a}\in \mathcal{A}}\boldsymbol{\pi}_{GA}^{\ast}(\boldsymbol{a} | s)\sum_{s'\in \mathcal{S}}P(s' | s, \boldsymbol{a})\left(V^{\pi_1^{\dagger}, \pi_{GA, 2}^{\ast}}(s') - V^{\boldsymbol{\pi}_{GA}^{\ast}}(s')\right) \\
    &\leq 2\delta + \gamma \max_{s'\in \mathcal{S}}\left(V^{\pi_1^{\dagger}, \pi_{GA, 2}^{\ast}}(s') - V^{\boldsymbol{\pi}_{GA}^{\ast}}(s')\right).
\end{align*}
Taking the maximum value of both sides, we have:
\begin{align*}
    \max_{s'\in \mathcal{S}} \left(V^{\pi_1^{\dagger}, \pi_{GA, 2}^{\ast}}(s') - V^{\boldsymbol{\pi}_{GA}^{\ast}}(s')\right) \leq 2\delta + \gamma \max_{s'\in \mathcal{S}}\left(V^{\pi_1^{\dagger}, \pi_{GA, 2}^{\ast}}(s') - V^{\boldsymbol{\pi}_{GA}^{\ast}}(s')\right).
\end{align*}
By moving $\gamma \max_{s'\in \mathcal{S}}(V^{\pi_1^{\dagger}, \pi_{GA, 2}^{\ast}}(s') - V^{\boldsymbol{\pi}_{GA}^{\ast}}(s'))$ from the right side to the left side, we have for any $s\in \mathcal{S}$:
\begin{align*}
    V^{\pi_1^{\dagger}, \pi_{GA, 2}^{\ast}}(s) - V^{\boldsymbol{\pi}_{GA}^{\ast}}(s) \leq \max_{s'\in \mathcal{S}} \left(V^{\pi_1^{\dagger}, \pi_{GA, 2}^{\ast}}(s') - V^{\boldsymbol{\pi}_{GA}^{\ast}}(s')\right) \leq \frac{2\delta}{1-\gamma}.
\end{align*}
\end{proof}

\subsection{Proof of Lemma \ref{lem:v_to_q}}
\label{app:v_to_q}
\begin{proof}[Proof of Lemma \ref{lem:v_to_q}]
From the definition of $V^{\pi_1^{\dagger}, \pi_{GA, 2}^{\ast}}$, we have for any $s\in \mathcal{S}$: 
\begin{align*}
    V^{\pi_1^{\dagger}, \pi_{GA, 2}^{\ast}}(s) = \sum_{\boldsymbol{a}\in \mathcal{A}}\pi_1^{\dagger}(a_1 | s)\pi_{GA, 2}^{\ast}(a_2 | s)Q^{\pi_1^{\dagger}, \pi_{GA, 2}^{\ast}}(s, \boldsymbol{a}).
\end{align*}
From the assumption of Lemma \ref{lem:v_to_q}, we get:
\begin{align*}
    V^{\pi_1^{\dagger}, \pi_{GA, 2}^{\ast}}(s) &\leq \sum_{\boldsymbol{a}\in \mathcal{A}}\pi_1^{\dagger}(a_1 | s)\pi_{GA, 2}^{\ast}(a_2 | s)Q_A^{\boldsymbol{\pi}_A^{\ast}}(\phi(s), \boldsymbol{a}) + \delta \\
    &= \sum_{\boldsymbol{a}\in \mathcal{A}}\pi_1^{\dagger}(a_1 | s)\pi_{A, 2}^{\ast}(a_2 | \phi(s))Q_A^{\boldsymbol{\pi}_A^{\ast}}(\phi(s), \boldsymbol{a}) + \delta \\
    &\leq \sum_{\boldsymbol{a}\in \mathcal{A}}\pi_{A,1}^{\ast}(a_1 | \phi(s))\pi_{A, 2}^{\ast}(a_2 | \phi(s))Q_A^{\boldsymbol{\pi}_A^{\ast}}(\phi(s), \boldsymbol{a}) + \delta \\
    &\leq \sum_{\boldsymbol{a}\in \mathcal{A}}\pi_{A,1}^{\ast}(a_1 | \phi(s))\pi_{A, 2}^{\ast}(a_2 | \phi(s))Q^{\pi_1^{\dagger}, \pi_{GA, 2}^{\ast}}(s, \boldsymbol{a}) + 2\delta \\
    &= \sum_{\boldsymbol{a}\in \mathcal{A}}\pi_{GA,1}^{\ast}(a_1 | s)\pi_{GA, 2}^{\ast}(a_2 | s)Q^{\pi_1^{\dagger}, \pi_{GA, 2}^{\ast}}(s, \boldsymbol{a}) + 2\delta.
\end{align*}
\end{proof}

% \subsection{Proof of Lemma \ref{lem:diff_between_v_and_v_abs_wrt_minimax_value}}
% \label{app:diff_between_v_and_v_abs_wrt_minimax_value}
% \begin{proof}[Proof of Lemma \ref{lem:diff_between_v_and_v_abs_wrt_minimax_value}]
% From the definition of $V_A^{\boldsymbol{\pi}_A^{\ast}}$, we have for any $s\in\mathcal{S}$:
% \begin{align*}
%     V_A^{\boldsymbol{\pi}_A^{\ast}}(\phi^{Q^{\ast}}(s)) &= \max_{p\in \Delta(A_1)}\min_{a_2\in \mathcal{A}_2} \sum_{a_1\in \mathcal{A}_1} p(a_1) Q_A^{\boldsymbol{\pi}_A^{\ast}}(\phi^{Q^{\ast}}(s), \boldsymbol{a}).
% \end{align*}
% Applying Lemma \ref{lem:diff_between_q_and_q_abs_wrt_minimax_value} to this equation, we get:
% \begin{align*}
%     & V_A^{\boldsymbol{\pi}_A^{\ast}}(\phi^{Q^{\ast}}(s)) \\
%     & \leq \max_{p\in \Delta(A_1)}\min_{a_2\in \mathcal{A}_2} \sum_{a_1\in \mathcal{A}_1} p(a_1) Q^{\ast}(s, \boldsymbol{a}) + \frac{\epsilon}{1-\gamma} \\
%     & = V^{\ast}(s) + \frac{\epsilon}{1-\gamma}.
% \end{align*}

% Similarly, 
% \begin{align*}
%     & V_A^{\boldsymbol{\pi}_A^{\ast}}(\phi^{Q^{\ast}}(s)) \\
%     & \geq \max_{p\in \Delta(A_1)}\min_{a_2\in \mathcal{A}_2} \sum_{a_1\in \mathcal{A}_1} p(a_1) Q^{\ast}(s, \boldsymbol{a}) - \frac{\epsilon}{1-\gamma} \\
%     &= V^{\ast}(s) - \frac{\epsilon}{1-\gamma}.
% \end{align*}

% Therefore, 
% \begin{align*}
%     \left|V^{\ast}(s) - V_A^{\boldsymbol{\pi}_A^{\ast}}(\phi^{Q^{\ast}}(s))\right| \leq \frac{\epsilon}{1-\gamma}.
% \end{align*}
% \end{proof}

% Model Similarity
\section{Proofs for Theorem \ref{thm:error_wrt_model_similarity}}
\label{sec:model_similarity_proof}

\subsection{Proof of Theorem~\ref{thm:error_wrt_model_similarity}}
\begin{proof}[Proof of Theorem \ref{thm:error_wrt_model_similarity}]
We introduce the following lemma:
\begin{lemma}
\label{lem:diff_between_q_and_q_abs}
In the same setup of Theorem \ref{thm:error_wrt_model_similarity}, we have for any $s\in \mathcal{S}$ and $\boldsymbol{a}\in \mathcal{A}$:
\begin{align*}
    \left|Q^{\pi_1^{\dagger}, \pi_{GA, 2}^{\ast}}(s, \boldsymbol{a}) - Q_A^{\boldsymbol{\pi}_A^{\ast}}(\phi^{\mathrm{model}}(s), \boldsymbol{a})\right| \leq  \frac{\epsilon\left(1 + \gamma \left(|\mathcal{S}|- 1\right)\right) }{(1-\gamma)^2}.
\end{align*}
\end{lemma}

By combining Lemma \ref{lem:exploitability_bound_minimax_value} and Lemma~\ref{lem:diff_between_q_and_q_abs}, we get for any $s\in \mathcal{S}$:
\begin{align*}
    V^{\pi_1^{\dagger}, \pi_{GA, 2}^{\ast}}(s) - V^{\boldsymbol{\pi}_{GA}^{\ast}}(s) \leq \frac{2\epsilon\left(1 + \gamma \left(|\mathcal{S}|- 1\right)\right) }{(1-\gamma)^2}.
\end{align*}
By a similar procedure, we can show that:
\begin{align*}
    V^{\boldsymbol{\pi}_{GA}^{\ast}}(s) - V^{\pi_{GA, 1}^{\ast}, \pi_2^{\dagger}}(s) \leq \frac{2\epsilon\left(1 + \gamma \left(|\mathcal{S}|- 1\right)\right) }{(1-\gamma)^2}.
\end{align*}
Therefore, from \eqref{eq:exploitability_of_pi_GA}, we obtain: 
\begin{align*}
    \mathrm{GAP}\left(\boldsymbol{\pi}_{GA}^{\ast}\right) \leq \frac{4(1 + \gamma \left(|\mathcal{S}|- 1\right))\epsilon}{(1-\gamma)^3}.
\end{align*}
\end{proof}

\subsection{Proof of Lemma \ref{lem:diff_between_q_and_q_abs}}
\begin{proof}
Firstly, from Assumption \ref{asmp:error_wrt_model_similarity}, if $\phi^{\mathrm{model}}(s_1)=\phi^{\mathrm{model}}(s_2)$, we have for any $\boldsymbol{a} \in \mathcal{A}$:
\begin{align}
    \left|Q^{\pi_1^{\dagger}, \pi_{GA, 2}^{\ast}}(s_1, \boldsymbol{a}) - Q^{\pi_1^{\dagger}, \pi_{GA, 2}^{\ast}}(s_2, \boldsymbol{a})\right| &\leq \left|R(s_1, \boldsymbol{a}) - R(s_2, \boldsymbol{a}) + \gamma \sum_{s'\in \mathcal{S}}\left(P(s'|s_1, \boldsymbol{a}) - P(s'|s_2, \boldsymbol{a})\right) V^{\pi_1^{\dagger}, \pi_{GA, 2}^{\ast}}(s')\right| \nonumber\\
    &\leq \left|R(s_1,\boldsymbol{a}) - R(s_2,\boldsymbol{a})\right| + \gamma \sum_{s'\in \mathcal{S}}\left|P(s'|s_1, \boldsymbol{a}) - P(s'|s_2, \boldsymbol{a})\right| \left|V^{\pi_1^{\dagger}, \pi_{GA, 2}^{\ast}}(s')\right| \nonumber\\
    &\leq \epsilon + \frac{\gamma \epsilon |\mathcal{S}|}{1-\gamma} = \frac{\epsilon + \gamma \left(|\mathcal{S}|- 1\right)\epsilon }{1-\gamma}.
    \label{eq:q_gap}
\end{align}

On the other hand, from the definition of the state-action value function, we have for $s\in \mathcal{S}$ and $\boldsymbol{a}\in \mathcal{A}$:
\begin{align*}
    Q_A^{\boldsymbol{\pi}_A^{\ast}}(\phi^{\mathrm{model}}(s), \boldsymbol{a}) &= \sum_{g\in G(s)} w(g) R(g, \boldsymbol{a}) + \gamma \sum_{g\in G(s)} w(g) \sum_{s'\in \mathcal{S}} P(s' | g, \boldsymbol{a}) \max_{a_1' \in \mathcal{A}_1}\sum_{a_2'\in \mathcal{A}_2} \pi_{GA, 2}^{\ast}(a_2' | s') Q_A^{\boldsymbol{\pi}_A^{\ast}}(\phi^{\mathrm{model}}(s'), \boldsymbol{a}') \\
    &\leq \sum_{g\in G(s)} w(g) R(g, \boldsymbol{a}) + \gamma \sum_{g\in G(s)} w(g) \sum_{s'\in \mathcal{S}} P(s' | g, \boldsymbol{a}) \max_{a_1' \in \mathcal{A}_1}\sum_{a_2'\in \mathcal{A}_2} \pi_{GA, 2}^{\ast}(a_2' | s') Q^{\pi_1^{\dagger}, \pi_{GA, 2}^{\ast}}(s', \boldsymbol{a}') \\
    &\phantom{\leq}+ \gamma \sum_{g\in G(s)} w(g) \sum_{s'\in \mathcal{S}} P(s' | g, \boldsymbol{a})\max_{(\boldsymbol{a}') \in \mathcal{A}} \left(Q_A^{\boldsymbol{\pi}_A^{\ast}}(\phi^{\mathrm{model}}(s'), \boldsymbol{a}') - Q^{\pi_1^{\dagger}, \pi_{GA, 2}^{\ast}}(s', \boldsymbol{a}')\right) \\
    &\leq \max_{g\in G(s)} Q^{\pi_1^{\dagger}, \pi_{GA, 2}^{\ast}}(g, \boldsymbol{a}) + \gamma \max_{(s', \boldsymbol{a}') \in \mathcal{S}\times \mathcal{A}} \left(Q_A^{\boldsymbol{\pi}_A^{\ast}}(\phi^{\mathrm{model}}(s'), \boldsymbol{a}') - Q^{\pi_1^{\dagger}, \pi_{GA, 2}^{\ast}}(s', \boldsymbol{a}')\right).
\end{align*}
Thus, from \eqref{eq:q_gap}, we have for $s\in \mathcal{S}$ and $\boldsymbol{a} \in \mathcal{A}$:
\begin{align*}
    &Q_A^{\boldsymbol{\pi}_A^{\ast}}(\phi^{\mathrm{model}}(s), \boldsymbol{a}) - Q^{\pi_1^{\dagger}, \pi_{GA, 2}^{\ast}}(s, \boldsymbol{a}) \\
    &\leq \max_{g\in G(s)} Q^{\pi_1^{\dagger}, \pi_{GA, 2}^{\ast}}(g, \boldsymbol{a}) - Q^{\pi_1^{\dagger}, \pi_{GA, 2}^{\ast}}(s, \boldsymbol{a}) + \gamma \max_{(s',\boldsymbol{a}') \in \mathcal{S}\times \mathcal{A}} \left(Q_A^{\boldsymbol{\pi}_A^{\ast}}(\phi^{\mathrm{model}}(s'), \boldsymbol{a}') - Q^{\pi_1^{\dagger}, \pi_{GA, 2}^{\ast}}(s', \boldsymbol{a}')\right) \\
    &\leq \frac{\epsilon + \gamma \left(|\mathcal{S}|- 1\right)\epsilon }{1-\gamma} + \gamma \max_{(s', \boldsymbol{a}') \in \mathcal{S}\times \mathcal{A}} \left(Q_A^{\boldsymbol{\pi}_A^{\ast}}(\phi^{\mathrm{model}}(s'), \boldsymbol{a}') - Q^{\pi_1^{\dagger}, \pi_{GA, 2}^{\ast}}(s', \boldsymbol{a}')\right).
\end{align*}
Taking the maximum of both sides:
\begin{align*}
    &\max_{(s', \boldsymbol{a}') \in \mathcal{S}\times \mathcal{A}} \left(Q_A^{\boldsymbol{\pi}_A^{\ast}}(\phi^{\mathrm{model}}(s'), \boldsymbol{a}') - Q^{\pi_1^{\dagger}, \pi_{GA, 2}^{\ast}}(s', \boldsymbol{a}')\right) \\
    &\leq \frac{\epsilon + \gamma \left(|\mathcal{S}|- 1\right)\epsilon }{1-\gamma} + \gamma \max_{(s', \boldsymbol{a}') \in \mathcal{S}\times \mathcal{A}} \left(Q_A^{\boldsymbol{\pi}_A^{\ast}}(\phi^{\mathrm{model}}(s'), \boldsymbol{a}') - Q^{\pi_1^{\dagger}, \pi_{GA, 2}^{\ast}}(s', \boldsymbol{a}')\right).
\end{align*}
Thus, we have for any $s\in \mathcal{S}$ and $\boldsymbol{a}\in \mathcal{A}$:
\begin{align*}
    Q_A^{\boldsymbol{\pi}_A^{\ast}}(\phi^{\mathrm{model}}(s), \boldsymbol{a}) - Q^{\pi_1^{\dagger}, \pi_{GA, 2}^{\ast}}(s, \boldsymbol{a}) &\leq \max_{(s', \boldsymbol{a}') \in \mathcal{S}\times \mathcal{A}} \left(Q_A^{\boldsymbol{\pi}_A^{\ast}}(\phi^{\mathrm{model}}(s'), \boldsymbol{a}') - Q^{\pi_1^{\dagger}, \pi_{GA, 2}^{\ast}}(s', \boldsymbol{a}')\right) \\
    &\leq \frac{\epsilon + \gamma \left(|\mathcal{S}|- 1\right)\epsilon }{(1-\gamma)^2}.
\end{align*}

Similarly, we have for any $s\in \mathcal{S}$ and $\boldsymbol{a}\in \mathcal{A}$:
\begin{align*}
    Q_A^{\boldsymbol{\pi}_A^{\ast}}(\phi^{\mathrm{model}}(s),\boldsymbol{a}) &\geq \sum_{g\in G(s)} w(g) R(g, \boldsymbol{a}) + \gamma \sum_{g\in G(s)} w(g) \sum_{s'\in \mathcal{S}} P(s' | g, \boldsymbol{a}) \max_{a_1' \in \mathcal{A}_1}\sum_{a_2'\in \mathcal{A}_2} \pi_{GA, 2}^{\ast}(a_2' | s') Q^{\pi_1^{\dagger}, \pi_{GA, 2}^{\ast}}(s', \boldsymbol{a}') \\
    &\phantom{\geq}+ \gamma \sum_{g\in G(s)} w(g) \sum_{s'\in \mathcal{S}} P(s' | g, \boldsymbol{a}) \min_{\boldsymbol{a}'\in \mathcal{A}} \left(Q_A^{\boldsymbol{\pi}_A^{\ast}}(\phi^{\mathrm{model}}(s'), \boldsymbol{a}') - Q^{\pi_1^{\dagger}, \pi_{GA, 2}^{\ast}}(s', \boldsymbol{a}')\right) \\
    &\geq \min_{g\in G(s)} Q^{\pi_1^{\dagger}, \pi_{GA, 2}^{\ast}}(g, \boldsymbol{a}) + \gamma \min_{(s', \boldsymbol{a}') \in \mathcal{S}\times \mathcal{A}} \left(Q_A^{\boldsymbol{\pi}_A^{\ast}}(\phi^{\mathrm{model}}(s'), \boldsymbol{a}') - Q^{\pi_1^{\dagger}, \pi_{GA, 2}^{\ast}}(s', \boldsymbol{a}')\right).
\end{align*}
Thus, from \eqref{eq:q_gap}, we have for any $s\in \mathcal{S}$ and $\boldsymbol{a} \in \mathcal{A}$:
\begin{align*}
    &Q_A^{\boldsymbol{\pi}_A^{\ast}}(\phi^{\mathrm{model}}(s),\boldsymbol{a}) - Q^{\pi_1^{\dagger}, \pi_{GA, 2}^{\ast}}(s,\boldsymbol{a}) \\
    &\geq \min_{g\in G(s)} Q^{\pi_1^{\dagger}, \pi_{GA, 2}^{\ast}}(g, \boldsymbol{a}) - Q^{\pi_1^{\dagger}, \pi_{GA, 2}^{\ast}}(s, \boldsymbol{a}) + \gamma \min_{(s',\boldsymbol{a}') \in \mathcal{S}\times \mathcal{A}} \left(Q_A^{\boldsymbol{\pi}_A^{\ast}}(\phi^{\mathrm{model}}(s'), \boldsymbol{a}') - Q^{\pi_1^{\dagger}, \pi_{GA, 2}^{\ast}}(s', \boldsymbol{a}')\right) \\
    &\geq -\frac{\epsilon + \gamma \left(|\mathcal{S}|- 1\right)\epsilon }{1-\gamma} + \gamma \min_{(s', \boldsymbol{a}') \in \mathcal{S}\times \mathcal{A}} \left(Q_A^{\boldsymbol{\pi}_A^{\ast}}(\phi^{\mathrm{model}}(s'), \boldsymbol{a}') - Q^{\pi_1^{\dagger}, \pi_{GA, 2}^{\ast}}(s', \boldsymbol{a}')\right).
\end{align*}
Taking the minimum of both sides:
\begin{align*}
    &\min_{(s',\boldsymbol{a}') \in \mathcal{S}\times \mathcal{A}} \left(Q_A^{\boldsymbol{\pi}_A^{\ast}}(\phi^{\mathrm{model}}(s'), \boldsymbol{a}') - Q^{\pi_1^{\dagger}, \pi_{GA, 2}^{\ast}}(s', \boldsymbol{a}')\right) \\
    &\geq -\frac{\epsilon + \gamma \left(|\mathcal{S}|- 1\right)\epsilon }{1-\gamma} + \gamma \min_{(s', \boldsymbol{a}') \in \mathcal{S}\times \mathcal{A}} \left(Q_A^{\boldsymbol{\pi}_A^{\ast}}(\phi^{\mathrm{model}}(s'), \boldsymbol{a}') - Q^{\pi_1^{\dagger}, \pi_{GA, 2}^{\ast}}(s', \boldsymbol{a}')\right).
\end{align*}
Hence, we have
\begin{align*}
    Q_A^{\boldsymbol{\pi}_A^{\ast}}(\phi^{\mathrm{model}}(s),\boldsymbol{a}) - Q^{\pi_1^{\dagger}, \pi_{GA, 2}^{\ast}}(s, \boldsymbol{a}) &\geq \min_{(s',\boldsymbol{a}') \in \mathcal{S}\times \mathcal{A}} \left(Q_A^{\boldsymbol{\pi}_A^{\ast}}(\phi^{\mathrm{model}}(s'), \boldsymbol{a}') - Q^{\pi_1^{\dagger}, \pi_{GA, 2}^{\ast}}(s', \boldsymbol{a}')\right) \\
    &\geq -\frac{\epsilon + \gamma \left(|\mathcal{S}|- 1\right)\epsilon }{(1-\gamma)^2}.
\end{align*}

In summary, we have for any $s\in\mathcal{S}$ and $\boldsymbol{a}\in\mathcal{A}$:
\begin{align*}
    \left|Q^{\pi_1^{\dagger}, \pi_{GA, 2}^{\ast}}(s, \boldsymbol{a}) - Q_A^{\boldsymbol{\pi}_A^{\ast}}(\phi^{\mathrm{model}}(s), \boldsymbol{a})\right| \leq  \frac{\epsilon + \gamma \left(|\mathcal{S}|- 1\right)\epsilon }{(1-\gamma)^2}.
\end{align*}
\end{proof}

% Boltzmann Distribution Similarity
\section{Proofs for Theorem \ref{thm:error_wrt_boltzmann_similarity}}
\label{sec:boltzmann_similarity_proof}

\subsection{Proof of Theorem \ref{thm:error_wrt_boltzmann_similarity}}
\begin{proof}
From Assumption \ref{asmp:error_wrt_boltzmann}, if $\phi^{\mathrm{bolt}}(s_1)=\phi^{\mathrm{bolt}}(s_2)$, we have for any $\boldsymbol{a}\in\mathcal{A}$:
\begin{align*}
    -\epsilon\sum_{\boldsymbol{b}\in \mathcal{A}}e^{Q^{\ast}(s_2, \boldsymbol{b})} &\leq \frac{e^{Q^{\ast}(s_1, \boldsymbol{a})}\sum_{\boldsymbol{b}\in \mathcal{A}}e^{Q^{\ast}(s_2, \boldsymbol{b})}}{\sum_{\boldsymbol{b}\in \mathcal{A}}e^{Q^{\ast}(s_1, \boldsymbol{b})}} - e^{Q^{\ast}(s_2, \boldsymbol{a})} \\
    &\leq \frac{e^{Q^{\ast}(s_1, \boldsymbol{a})}\left(\sum_{\boldsymbol{b}\in \mathcal{A}}e^{Q^{\ast}(s_1, \boldsymbol{b})} + k\epsilon\right)}{\sum_{\boldsymbol{b}\in \mathcal{A}}e^{Q^{\ast}(s_1, \boldsymbol{b})}} - e^{Q^{\ast}(s_2, \boldsymbol{a})} \\
    &= k\epsilon\frac{e^{Q^{\ast}(s_1, \boldsymbol{a})}}{\sum_{\boldsymbol{b}\in \mathcal{A}}e^{Q^{\ast}(s_1, \boldsymbol{b})}} + e^{Q^{\ast}(s_1, \boldsymbol{a})} - e^{Q^{\ast}(s_2, \boldsymbol{a})}.
\end{align*}
Similarly, if $\phi^{\mathrm{bolt}}(s_1)=\phi^{\mathrm{bolt}}(s_2)$, we have:
\begin{align*}
    \epsilon\sum_{\boldsymbol{b}\in \mathcal{A}}e^{Q^{\ast}(s_2, \boldsymbol{b})} &\geq \frac{e^{Q^{\ast}(s_1, \boldsymbol{a})}\sum_{\boldsymbol{b}\in \mathcal{A}}e^{Q^{\ast}(s_2, \boldsymbol{b})}}{\sum_{\boldsymbol{b}\in \mathcal{A}}e^{Q^{\ast}(s_1, \boldsymbol{b})}} - e^{Q^{\ast}(s_2, \boldsymbol{a})} \\
    &\geq \frac{e^{Q^{\ast}(s_1, \boldsymbol{a})}\left(\sum_{\boldsymbol{b}\in \mathcal{A}}e^{Q^{\ast}(s_1, \boldsymbol{b})} - k\epsilon\right)}{\sum_{\boldsymbol{b}\in \mathcal{A}}e^{Q^{\ast}(s_1, \boldsymbol{b})}} - e^{Q^{\ast}(s_2, \boldsymbol{a})} \\
    &= -k\epsilon\frac{e^{Q^{\ast}(s_1, \boldsymbol{a})}}{\sum_{\boldsymbol{b}\in \mathcal{A}}e^{Q^{\ast}(s_1, \boldsymbol{b})}} + e^{Q^{\ast}(s_1, \boldsymbol{a})} - e^{Q^{\ast}(s_2, \boldsymbol{a})}.
\end{align*}
Hence, we obtain for any $\boldsymbol{a}\in \mathcal{A}$ and $s_1, s_2\in \mathcal{S}$ such that $\phi^{\mathrm{bolt}}(s_1)=\phi^{\mathrm{bolt}}(s_2)$:
\begin{align}
    \left|e^{Q^{\ast}(s_1, \boldsymbol{a})} - e^{Q^{\ast}(s_2, \boldsymbol{a})}\right| \leq \epsilon \left(\sum_{\boldsymbol{b}\in \mathcal{A}}e^{Q^{\ast}(s_2, \boldsymbol{b})} + k\frac{e^{Q^{\ast}(s_1, \boldsymbol{a})}}{\sum_{\boldsymbol{b}\in \mathcal{A}}e^{Q^{\ast}(s_1, \boldsymbol{b})}}\right).
    \label{eq:diff_exponential_minimax_value}
\end{align}

On the other hand, from the mean value theorem, for any $x, y\in [-\frac{1}{1-\gamma}, \frac{1}{1 - \gamma}]$, there exists some constant $z \in [-\frac{1}{1-\gamma}, \frac{1}{1 - \gamma}]$ such that:
\begin{align*}
    \frac{e^x - e^y}{x - y} = e^z.
\end{align*}
Thus, we get:
\begin{align}
    \left|x - y\right| = \frac{\left|e^x - e^y\right|}{e^z} \leq e^{\frac{1}{1-\gamma}}\left|e^x - e^y\right|.
    \label{eq:diff_exponential_fn}
\end{align}

By combining \eqref{eq:diff_exponential_minimax_value} and \eqref{eq:diff_exponential_fn}, we have for any $s\in \mathcal{S}$ and $\boldsymbol{a}\in \mathcal{A}$:
\begin{align*}
    \left|Q^{\ast}(s_1, \boldsymbol{a}) - Q^{\ast}(s_2, \boldsymbol{a})\right| &\leq \epsilon e^{\frac{1}{1-\gamma}} \left(\sum_{\boldsymbol{b}\in \mathcal{A}}e^{Q^{\ast}(s_2, \boldsymbol{b})} + k\frac{e^{Q^{\ast}(s_1, \boldsymbol{a})}}{\sum_{\boldsymbol{b}\in \mathcal{A}}e^{Q^{\ast}(s_1, \boldsymbol{b})}}\right) \\
    &\leq \epsilon \left(|\mathcal{A}_1||\mathcal{A}_2|e^{\frac{2}{1-\gamma}} + k\frac{e^{\frac{3}{1-\gamma}}}{|\mathcal{A}_1||\mathcal{A}_2|}\right) \\
    &= e^{\frac{2}{1-\gamma}}\left(|\mathcal{A}_1||\mathcal{A}_2| + k\frac{e^{\frac{1}{1-\gamma}}}{|\mathcal{A}_1||\mathcal{A}_2|}\right) \epsilon.
\end{align*}
This inequality implies that, under the aggregation function $\phi^{\mathrm{bolt}}$, Assumption \ref{asmp:minimax_Q_abstraction} holds with $e^{\frac{2}{1-\gamma}}\left(|\mathcal{A}_1||\mathcal{A}_2| + k\frac{e^{\frac{1}{1-\gamma}}}{|\mathcal{A}_1||\mathcal{A}_2|}\right) \epsilon$.
Therefore, we can apply Theorem \ref{thm:error_wrt_minimax_value}, leading that:
\begin{align*}
    \mathrm{GAP}\left(\boldsymbol{\pi}_{GA}^{\ast}\right) \leq \frac{12e^{\frac{2}{1-\gamma}}\left(|\mathcal{A}_1||\mathcal{A}_2| + k\frac{e^{\frac{1}{1-\gamma}}}{|\mathcal{A}_1||\mathcal{A}_2|}\right) \epsilon}{(1-\gamma)^3}.
\end{align*}
\end{proof}

% Multinomial Distribution Similartity
\section{Proofs for Theorem \ref{thm:error_wrt_multinomial}}
\label{sec:multinomial_similarity_proof}

\subsection{Proof of Theorem \ref{thm:error_wrt_multinomial}}

\begin{proof}
From Assumption \ref{asmp:error_wrt_multinomial}, if $\sum_{\boldsymbol{b}\in \mathcal{A}}Q^{\ast}(s_2, \boldsymbol{b}) \geq 0$, then we have for any $\boldsymbol{a}\in \mathcal{A}$ and $s_1, s_2\in \mathcal{S}$ such that $\phi^{\mathrm{mult}}(s_1)=\phi^{\mathrm{mult}}(s_2)$:
\begin{align*}
    -\epsilon\sum_{\boldsymbol{b}\in \mathcal{A}}Q^{\ast}(s_2, \boldsymbol{b}) &\leq \frac{Q^{\ast}(s_1, \boldsymbol{a})\sum_{\boldsymbol{b}\in \mathcal{A}}Q^{\ast}(s_2, \boldsymbol{b})}{\sum_{\boldsymbol{b}\in \mathcal{A}}Q^{\ast}(s_1, \boldsymbol{b})} - Q^{\ast}(s_2, \boldsymbol{a}) \\
    &\leq \frac{Q^{\ast}(s_1, \boldsymbol{a})\left(\sum_{\boldsymbol{b}\in \mathcal{A}}Q^{\ast}(s_1, \boldsymbol{b}) + k\epsilon\right)}{\sum_{\boldsymbol{b}\in \mathcal{A}}Q^{\ast}(s_1, \boldsymbol{b})} - Q^{\ast}(s_2, \boldsymbol{a}) \\
    &= k\epsilon\frac{Q^{\ast}(s_1, \boldsymbol{a})}{\sum_{\boldsymbol{b}\in \mathcal{A}}Q^{\ast}(s_1, \boldsymbol{b})} + Q^{\ast}(s_1, \boldsymbol{a}) - Q^{\ast}(s_2, \boldsymbol{a}).
\end{align*}
Similarly, we have for any $\boldsymbol{a}\in \mathcal{A}$ and $s_1, s_2\in \mathcal{S}$ such that $\phi^{\mathrm{mult}}(s_1)=\phi^{\mathrm{mult}}(s_2)$:
\begin{align*}
    \epsilon\sum_{\boldsymbol{b}\in \mathcal{A}}Q^{\ast}(s_2, \boldsymbol{b}) &\geq \frac{Q^{\ast}(s_1, \boldsymbol{a})\sum_{\boldsymbol{b}\in \mathcal{A}}Q^{\ast}(s_2, \boldsymbol{b})}{\sum_{\boldsymbol{b}\in \mathcal{A}}Q^{\ast}(s_1, \boldsymbol{b})} - Q^{\ast}(s_2, \boldsymbol{a}) \\
    &\geq \frac{Q^{\ast}(s_1, \boldsymbol{a})\left(\sum_{\boldsymbol{b}\in \mathcal{A}}Q^{\ast}(s_1, \boldsymbol{b}) - k\epsilon\right)}{\sum_{\boldsymbol{b}\in \mathcal{A}}Q^{\ast}(s_1, \boldsymbol{b})} - Q^{\ast}(s_2, \boldsymbol{a}) \\
    &= -k\epsilon\frac{Q^{\ast}(s_1, \boldsymbol{a})}{\sum_{\boldsymbol{b}\in \mathcal{A}}Q^{\ast}(s_1, \boldsymbol{b})} + Q^{\ast}(s_1, \boldsymbol{a}) - Q^{\ast}(s_2, \boldsymbol{a}).
\end{align*}
Hence, if $\sum_{\boldsymbol{b}\in \mathcal{A}}Q^{\ast}(s_2, \boldsymbol{b}) \geq 0$, we obtain:
\begin{align}
    \left|Q^{\ast}(s_1, \boldsymbol{a}) - Q^{\ast}(s_2, \boldsymbol{a})\right| \leq \epsilon \left(\sum_{\boldsymbol{b}\in \mathcal{A}}Q^{\ast}(s_2, \boldsymbol{b}) + k\frac{Q^{\ast}(s_1, \boldsymbol{a})}{\sum_{\boldsymbol{b}\in \mathcal{A}}Q^{\ast}(s_1, \boldsymbol{b})}\right).
    \label{eq:diff_exponential_minimax_value_wrt_multinomial_1}
\end{align}

On the other hand, if $\sum_{\boldsymbol{b}\in \mathcal{A}}Q^{\ast}(s_2, \boldsymbol{b}) < 0$, then we have for any $\boldsymbol{a}\in \mathcal{A}$ and $s_1, s_2\in \mathcal{S}$ such that $\phi^{\mathrm{mult}}(s_1)=\phi^{\mathrm{mult}}(s_2)$:
\begin{align*}
    -\epsilon\sum_{\boldsymbol{b}\in \mathcal{A}}Q^{\ast}(s_2, \boldsymbol{b}) &\geq \frac{Q^{\ast}(s_1, \boldsymbol{a})\sum_{\boldsymbol{b}\in \mathcal{A}}Q^{\ast}(s_2, \boldsymbol{b})}{\sum_{\boldsymbol{b}\in \mathcal{A}}Q^{\ast}(s_1, \boldsymbol{b})} - Q^{\ast}(s_2, \boldsymbol{a}) \\
    &\geq \frac{Q^{\ast}(s_1, \boldsymbol{a})\left(\sum_{\boldsymbol{b}\in \mathcal{A}}Q^{\ast}(s_1, \boldsymbol{b}) - k\epsilon\right)}{\sum_{\boldsymbol{b}\in \mathcal{A}}Q^{\ast}(s_1, \boldsymbol{b})} - Q^{\ast}(s_2, \boldsymbol{a}) \\
    &= -k\epsilon\frac{Q^{\ast}(s_1, \boldsymbol{a})}{\sum_{\boldsymbol{b}\in \mathcal{A}}Q^{\ast}(s_1, \boldsymbol{b})} + Q^{\ast}(s_1, \boldsymbol{a}) - Q^{\ast}(s_2, \boldsymbol{a}).
\end{align*}
Similarly, we have:
\begin{align*}
    \epsilon\sum_{\boldsymbol{b}\in \mathcal{A}}Q^{\ast}(s_2, \boldsymbol{b}) &\leq \frac{Q^{\ast}(s_1, \boldsymbol{a})\sum_{\boldsymbol{b}\in \mathcal{A}}Q^{\ast}(s_2, \boldsymbol{b})}{\sum_{\boldsymbol{b}\in \mathcal{A}}Q^{\ast}(s_1, \boldsymbol{b})} - Q^{\ast}(s_2, \boldsymbol{a}) \\
    &\leq \frac{Q^{\ast}(s_1, \boldsymbol{a})\left(\sum_{\boldsymbol{b}\in \mathcal{A}}Q^{\ast}(s_1, \boldsymbol{b}) + k\epsilon\right)}{\sum_{\boldsymbol{b}\in \mathcal{A}}Q^{\ast}(s_1, \boldsymbol{b})} - Q^{\ast}(s_2, \boldsymbol{a}) \\
    &= k\epsilon\frac{Q^{\ast}(s_1, \boldsymbol{a})}{\sum_{\boldsymbol{b}\in \mathcal{A}}Q^{\ast}(s_1, \boldsymbol{b})} + Q^{\ast}(s_1, \boldsymbol{a}) - Q^{\ast}(s_2, \boldsymbol{a}).
\end{align*}
Thus, if $\sum_{\boldsymbol{b}\in \mathcal{A}}Q^{\ast}(s_2, \boldsymbol{b}) < 0$, we get:
\begin{align}
    \left|Q^{\ast}(s_1, \boldsymbol{a}) - Q^{\ast}(s_2, \boldsymbol{a})\right| \leq \epsilon \left(-\sum_{\boldsymbol{b}\in \mathcal{A}}Q^{\ast}(s_2, \boldsymbol{b}) + k\frac{Q^{\ast}(s_1, \boldsymbol{a})}{\sum_{\boldsymbol{b}\in \mathcal{A}}Q^{\ast}(s_1, \boldsymbol{b})}\right).
    \label{eq:diff_exponential_minimax_value_wrt_multinomial_2}
\end{align}

By combining \eqref{eq:diff_exponential_minimax_value_wrt_multinomial_1} and \eqref{eq:diff_exponential_minimax_value_wrt_multinomial_2}, we have for any $\boldsymbol{a}\in \mathcal{A}$ and $s_1, s_2\in \mathcal{S}$ such that $\phi^{\mathrm{mult}}(s_1)=\phi^{\mathrm{mult}}(s_2)$:
\begin{align*}
    \left|Q^{\ast}(s_1, \boldsymbol{a}) - Q^{\ast}(s_2, \boldsymbol{a})\right| &\leq \epsilon \left(\left|\sum_{\boldsymbol{b}\in \mathcal{A}}Q^{\ast}(s_2, \boldsymbol{b})\right| + k\frac{Q^{\ast}(s_1, \boldsymbol{a})}{\sum_{\boldsymbol{b}\in \mathcal{A}}Q^{\ast}(s_1, \boldsymbol{b})}\right) \\
    &\leq \epsilon \left(\frac{|\mathcal{A}_1||\mathcal{A}_2|}{1-\gamma} + \frac{k}{(1-\gamma)\delta}\right) \\
    &= \frac{1}{1-\gamma} \left(|\mathcal{A}_1||\mathcal{A}_2| + \frac{k}{\delta}\right)\epsilon.
\end{align*}
This inequality implies that, under the aggregation function $\phi^{\mathrm{mult}}$, Assumption \ref{asmp:minimax_Q_abstraction} holds with $\frac{1}{1-\gamma} \left(|\mathcal{A}_1||\mathcal{A}_2| + \frac{k}{\delta}\right)\epsilon$.
Therefore, we can apply Theorem \ref{thm:error_wrt_minimax_value}, leading that
\begin{align*}
    \mathrm{GAP}\left(\boldsymbol{\pi}_{GA}^{\ast}\right) \leq \frac{12\left(|\mathcal{A}_1||\mathcal{A}_2| + \frac{k}{\delta}\right)\epsilon}{(1-\gamma)^4}.
\end{align*}
\end{proof}
    
% \end{comment}

\end{document}